\documentclass[sigconf]{acmart}

\copyrightyear{2024}
\acmYear{2024}
\setcopyright{acmlicensed}\acmConference[CCS '24]{Proceedings of the 2024
ACM SIGSAC Conference on Computer and Communications Security}{October
14--18, 2024}{Salt Lake City, UT, USA}
\acmBooktitle{Proceedings of the 2024 ACM SIGSAC Conference on Computer
and Communications Security (CCS '24), October 14--18, 2024, Salt Lake City,
UT, USA}
\acmDOI{10.1145/3658644.3690345}
\acmISBN{979-8-4007-0636-3/24/10}

\usepackage{microtype}
\usepackage[subtle]{savetrees}
\usepackage{pifont}
\usepackage{tcolorbox}
\usepackage{multicol}
\usepackage{multirow}
\usepackage[toc,page]{appendix}

\newtheorem{definition}{Definition}
\usepackage{bbm}
\usepackage{comment}
\usepackage{graphicx}
\usepackage{subfigure}
\usepackage{url}
\usepackage{amsmath}
\usepackage{caption}
\usepackage{mathtools}
\usepackage{enumitem}
\usepackage{booktabs}
\usepackage{amsmath}
\usepackage{tabularray}
\usepackage{tabularx}
\usepackage{tablefootnote}
\UseTblrLibrary{booktabs}

\newtheorem{requirement}{Rule}
\usepackage{tcolorbox}
\usepackage{algorithmic}
\usepackage{float}
\usepackage{algorithm}

\newcommand{\projname}{{\large\textsc{Marionette}}}
\newcommand{\projnameTable}{{\textsc{Marionette}}}

\newcommand{\kw}[1]{\texttt{#1}}
\newcommand{\switch}[1]{\textit{#1}}
\newcommand{\inlinedsection}[2][3pt]{\vspace{#1}\noindent\textbf{#2}.} 
\newcommand{\inlinedsectionit}[2][3pt]{\vspace{#1}\noindent\textit{#2}:} 

\newcommand{\attack}[1]{\textbf{(A.#1)}}
\newcommand{\defense}[1]{\textbf{(D.#1)}}

\newif\ifsubmit
\submittrue 
\ifsubmit
\newcommand{\mm}[1]{}
\newcommand{\teryl}[1]{}
\newcommand{\attention}[1]{}
\newcommand{\fred}[1]{}
\newcommand{\tpt}[1]{} 
\newcommand{\tom}[1]{}
\newcommand{\trent}[1]{}
\newcommand{\todo}[1]{}
\newcommand{\inspiration}[1]{}
\newcommand{\update}[1]{#1}
\else
\definecolor{darkviolet}{rgb}{0.58, 0.0, 0.83}
\newcommand{\mm}[1]{\textcolor{darkviolet}{\textbf{Mingming:} #1}}
\newcommand{\teryl}[1]{\textcolor{cyan}{Teryl: #1}}
\newcommand{\attention}[1]{\textcolor{red}{ATTENTION: #1}}
\definecolor{crimsonglory}{rgb}{0.75, 0.0, 0.2}
\newcommand{\fred}[1]{\textcolor{crimsonglory}{\textbf{Fred:} #1}}
\newcommand{\tpt}[1]{\textcolor{red}{Teryl: #1}} 
\definecolor{crimson}{rgb}{0.86, 0.08, 0.24}
\newcommand{\tom}[1]{\textcolor{crimson}{\textbf{Tom:} #1}}
\definecolor{alizarin}{rgb}{0.82, 0.1, 0.26}
\newcommand{\trent}[1]{\textcolor{alizarin}{\textbf{Trent:} #1}}
\newcommand{\todo}[1]{\textcolor{red}{\textbf{TODO:} #1}}
\newcommand{\inspiration}[1]{\textcolor{blue}{\textbf{Inspiration:} #1}}
\newcommand{\update}[1]{\textcolor{blue}{#1}}
\fi

\begin{document}

\title{Manipulating OpenFlow Link Discovery Packet Forwarding for Topology Poisoning}

\advance\baselineskip-.32pt plus.2pt minus.3pt

\author{Mingming Chen}
\email{mzc796@psu.edu}
\orcid{0000-0001-9595-770X}
\affiliation{%
  \institution{The Pennsylvania State University}
  \city{University Park}
  \state{PA}
  \country{USA}
}

\author{Thomas La Porta}
\email{tfl12@psu.edu}
\affiliation{%
  \institution{The Pennsylvania State University}
  \city{University Park}
  \state{PA}
  \country{USA}
}

\author{Teryl Taylor}
\email{terylt@ibm.com}
\affiliation{%
  \institution{IBM Research}
  \city{Yorktown Heights}
  \state{NY}
  \country{USA}
}

\author{Frederico Araujo}
\email{frederico.araujo@ibm.com}
\affiliation{%
  \institution{IBM Research}
  \city{Yorktown Heights}
  \state{NY}
  \country{USA}
}

\author{Trent Jaeger}
\email{trentj@ucr.edu}
\orcid{0000-0002-4964-1170}
\affiliation{%
  \institution{University of California, Riverside}
  \city{Riverside}
  \state{CA}
  \country{USA}
}

\begin{abstract}
Software-defined networking (SDN) is a centralized, dynamic, and programmable network management technology that enables flexible traffic control and scalability. SDN facilitates network administration through a centralized view of the underlying physical topology; tampering with this topology view can result in catastrophic damage to network management and security. 
To underscore this issue, we introduce  \projname{}, a new topology poisoning technique that manipulates OpenFlow link discovery packet forwarding to alter topology information. Our approach exposes an overlooked yet widespread attack vector, distinguishing itself from traditional link fabrication attacks that tamper, spoof, or relay discovery packets at the data plane. 
Unlike localized attacks observed in existing methods, our technique introduces a globalized topology poisoning attack that leverages control privileges. \projname{} implements a reinforcement learning algorithm to compute a poisoned topology target, and injects flow entries to achieve a long-lived stealthy attack.
Our evaluation shows that \projname{} successfully attacks five open-source controllers and nine OpenFlow-based discovery protocols. \projname{} overcomes the state-of-the-art topology poisoning defenses, showcasing a new class of topology poisoning that initiates on the control plane. This security vulnerability was ethically disclosed to OpenDaylight, and CVE-2024-37018 has been assigned.   
\end{abstract}

\settopmatter{printfolios=true}

\maketitle
\section{Introduction}
Software-defined networking (SDN) is a flexible network architecture that offers centralized control and management of a network, in contrast to traditional data networks. SDN enables a controller to manage the network centrally through one of several protocols (e.g., OpenFlow~\cite{openflow}) and decouples control from the switches.  
A major advantage of SDN is that it provides a real-time view of the entire network, which allows for flexible traffic management. This serves as the foundation for many services, including cloud computing, traffic engineering, and network monitoring. 

Despite its many advantages, SDN's centralized architecture suffers from inherent limitations related to scalability and fault-tolerance~\cite{ahmad2015security,zhang2018survey}.
To solve these issues, multi-controller architectures were introduced~\cite{zhang2018survey, hu2018multi} to reduce communication latency between controllers and switches and balance the load across multiple controllers. Furthermore, fault-tolerant controller clustering enables rapid recovery of the control plane to mitigate the impact of a single point of failure or attacks on a single controller. 
However, multi-controller architectures also introduce attack vectors~\cite{maleh2023comprehensive,shaghaghi2020software,abdou2018comparative,kreutz2014software} spanning the \emph{east-westbound} interface (between controllers) and \emph{north-southbound} interface (between controllers and applications/switches).

Specifically, while topology poisoning attacks originating from the data plane have been well-studied~\cite{hong2015poisoning,nguyen2017analysis,alimohammadifar2018stealthy}, the control plane 
poses challenges for creating sophisticated, long-lasting topology poisoning attacks. 
\update{
For example, a na\"ive approach, in which a malicious controller 
shares incorrect information with peer controllers to poison the topology,
is not effective~\cite{khan2016topology} because leader controllers periodically (e.g., every 100 milliseconds~\cite{wazirali2021sdn}) re-discover the topology directly from the network, independently~\cite{wazirali2021sdn,azzouni2017softdp,dhawan2015sphinx}. Moreover, the assignment of the leader role is dynamic~\cite{hu2018multi,zhang2018survey,tootoonchian2010hyperflow}, rendering attacks that rely on a static leader ineffective. Similarly,  attacks 
that \emph{directly} manipulate 
traffic forwarding through the injection of malicious flow entries 
are easily detectable~\cite{Gwardar18, ropke2018preventing, lee2018indago, matsumoto2014fleet}.}

\update{A critical and previously overlooked vulnerability in modern SDN link discovery occurs when the flow entries designed for traffic routing are used to impact link discovery. 
The OpenFlow Discovery Protocol (OFDP), the de-facto SDN discovery protocol,  enables controllers to perform link discovery between any pair of {neighboring} switches; however, maliciously constructed (i.e., poisonous\footnote{~We use ``poisonous" as the adjective of flow entries that poison the topology. Consequently, such topology is called a ``poisoned topology".}
) flow entries that manipulate the Link Layer Discovery Protocol (LLDP)\footnote{~A link layer protocol used by IEEE 802 network devices to learn reachability and connection endpoint information from adjacent devices~\cite{lldp}.} packet forwarding may cause controllers to discover incorrect links and miss real ones. 
Further, we find that both the OpenDaylight and ONOS controller clusters\footnote{~The only two Open-Source controller projects with clustering implementation.}
allow controllers in any role in a multi-controller architecture to introduce these poisonous flow entries due to an insecure multi-controller implementation. Malicious applications on an SDN controller can also exploit this link discovery vulnerability. 
} 

To underscore the severity of this security vulnerability, we introduce \projname{}, a \emph{persistent}, \emph{stealthy}, \emph{precise}, and \emph{globalized} topology poisoning attack that uses poisonous OpenFlow~\cite{openflow} flow entries to induce controllers to independently discover a poisoned topology using standard discovery protocols. 
In contrast to 
techniques that ephemerally mislead controllers with false information, our approach steers benign controllers to independently discover a false, poisoned topology and accept it indefinitely,  without flow interruption.
\update{Modern topology discovery protocols infer links between switches based on the start and end points of the path traversed by an LLDP packet. \projname{} manipulates the traversal's endpoint to fabricate links, circumventing current state-of-the-art detection mechanisms.}
This enables \projname{} to manipulate any discoverable links in the network irrespective of their location.
As a result, it can initiate a global attack by computing a target poisoned topology based on the real topology.

\projname{} works with standard OpenFlow protocols, attacking OpenFlow-based SDN topology discovery protocols that flood or broadcast discovery packets to discover links (\S\ref{sec:eval_protocol}). 
The attack first learns the current network topology and forwarding policy.
It then uses reinforcement learning (RL)~\cite{russell2016artificial} to learn a deceptive topology that will achieve an attack goal (e.g., evading a monitor, attracting traffic to an eavesdropping switch).
When building the RL model, constraints are imposed to evade existing defenses~\cite{dhawan2015sphinx} while limiting the differences between the poisoned and real topologies (\S\ref{sec:design_RL}). Once a target topology is determined, \projname{} automatically derives the required poisonous flow entries and sends them to the appropriate switches concealed as normal flow entries (\S\ref{sec:simple_pois_entry}, \S\ref{sec:vlanentries}). Finally, while gaps in routes---introduced by differences between the real and fabricated topologies---can be fixed through reactive forwarding~\cite{openflow}, we use 
a stealthier approach to patch these differences via additional 
flow entries that minimize suspicious \kw{packet-in} packets sent to the benign controller (\S\ref{sec:gappatchentries}). 
Performing these three functions 
concurrently in a real network is not trivial. The end-to-end attack strategy is discussed in \S\ref{sec:design}.

\projname{} is robust against encryption-based defenses~\cite{azzouni2017softdp,hong2015poisoning,alharbi2015security,adjou2022topotrust} because it does not rely on packet spoofing or fabrication. Software rootkit defenses~\cite{ropke2018preventing} are also ineffective since the attack only relies on normal controller functionality. Moreover, the poisonous flow entries generated by \projname{} escape flow rule examinations~\cite{khurshid2012veriflow,porras2012security} and network policy checking~\cite{kazemian2013real} (\S\ref{sec:eval_defense}).  \update{\projname{} overcomes scalability issues by
computing its deceptive topologies offline and performing infrequent topology changes. }

To validate the practicality of our approach, we implemented \projname{} attacks against {nine} OpenFlow-based discovery protocols and {five} SDN controllers, including attacks against OpenDaylight and ONOS \emph{clusters}.  
We also used \projname{}'s RL algorithm to compute deceptive topologies for two network topologies, \textit{fat tree}~\cite{Leiserson:1985}, which is widely used in enterprise networks, and \textit{Chinanet}~\cite{topozoo}, a backbone topology. \update{Our experiments show that \projname{} 
generates a poisoned topology that is 92\% similar to the original topology while attracting more than 60\% additional flows to the eavesdropping point on a 36-node fat tree topology. We analyzed current state-of-the-art defenses against \projname{}, revealing that all are ineffective against our attack.} 
\update{We have ethically disclosed our attack details to the SDN controllers we evaluated, including OpenDayLight, which has acknowledged the vulnerability.}

Our contributions can be summarized as follows:

\begin{itemize}[itemsep=1pt, topsep=2pt, leftmargin=12pt]
    \item We introduce a novel method for topology poisoning initiated from the control plane which generates poisonous flow entries that manipulate link discovery packets' forwarding to precisely fabricate links toward a specific attack goal. 
    
    \item We design a reinforcement learning algorithm that automatically computes network topologies that satisfy attack objectives (e.g., flow routing through an attacker-controlled node) and constraints (e.g., the deceptive topology has a certain graph similarity with the real topology).
    
    \item We successfully deployed \projname{} attacks on both ONOS and OpenDaylight clusters, and five open-source controllers, \update{while systematically evading existing defenses.} We also show that \projname{} can attack nine different SDN discovery protocols, and evaluate our RL model on two real topologies. 
\end{itemize}

\section{Overview}\label{sec:overview_motivation}

\begin{table}[t]
\caption{Control Plane Vulnerabilities}
\footnotesize
\label{table:vulnerability}
\begin{tabular}{@{}>{\raggedright}p{3.2cm} p{4.9cm}@{}}
    \toprule
    \textbf{Vulnerabilities} & \textbf{Security issues~|~CVEs}\\
    \midrule
    Malicious Controller & \cite{khan2016topology},\cite{matsumoto2014fleet},\cite{qi2016intensive},\cite{zhou2018sdn},\cite{adjou2022topotrust},\cite{li2014byzantine}\\
    \midrule
    Controller Impersonation & \cite{maleh2023comprehensive},\cite{ kreutz2014software},\cite{onf_threat_analysis}\cite{tseng2018comprehensive}, Appendix \ref{sec:eval_impersonation}\\
    \midrule
    
    Unauthenticated Access & \cite{unauthenticated_access_cisco},\cite{chang2018fast},\cite{maleh2023comprehensive} \\
    \midrule
    
    Controller Vulnerabilities& CVE-2018-1000614, CVE-2023-30093, CVE-2017-1000081, CVE-2018-1132, CVE-2017-1000406\\
    \midrule
    Malicious Application & \cite{al2010flowchecker},\cite{habib2022mitigating},\cite{ujcich2018cross},\cite{zhou2018sdn},\cite{ujcich2021causal},\cite{ropke2018preventing},\cite{maleh2023comprehensive},\cite{cao2020match}\\
    \midrule
    Northbound API abuse & \cite{hu2021seapp},\cite{tseng2017controller}\\
    \bottomrule
\end{tabular}
\end{table}

SDN provides dynamic, flexible, and programmable traffic management by decoupling the control plane from the data plane. Unlike legacy networks running a distributed routing algorithm on each switch, SDN employs a centralized controller to gather network information and orchestrate actions across switches. To enhance fault tolerance and scalability of the single-controller architecture, a multi-controller architecture was developed~\cite{zhang2018survey,hu2018multi}.
However, the increased complexity and multiple interfaces expose many types of vulnerabilities, especially on the control plane, as listed in Table \ref{table:vulnerability}.

\update{A malicious controller in a cluster or malicious application in the SDN platform can jeopardize the network by proactively injecting malicious flow entries. As a result, defenses have been developed to detect malicious flow entries that cause routing attacks~\cite{khurshid2012veriflow, al2010flowchecker,porras2012security};
however, a heretofore overlooked attack inserts poisonous flow entries that \emph{manipulate link discovery packet forwarding} to poison the topology view of benign controllers, evading existing defenses.}

\begin{figure}[t]
\centering{\includegraphics[width=0.45\textwidth]{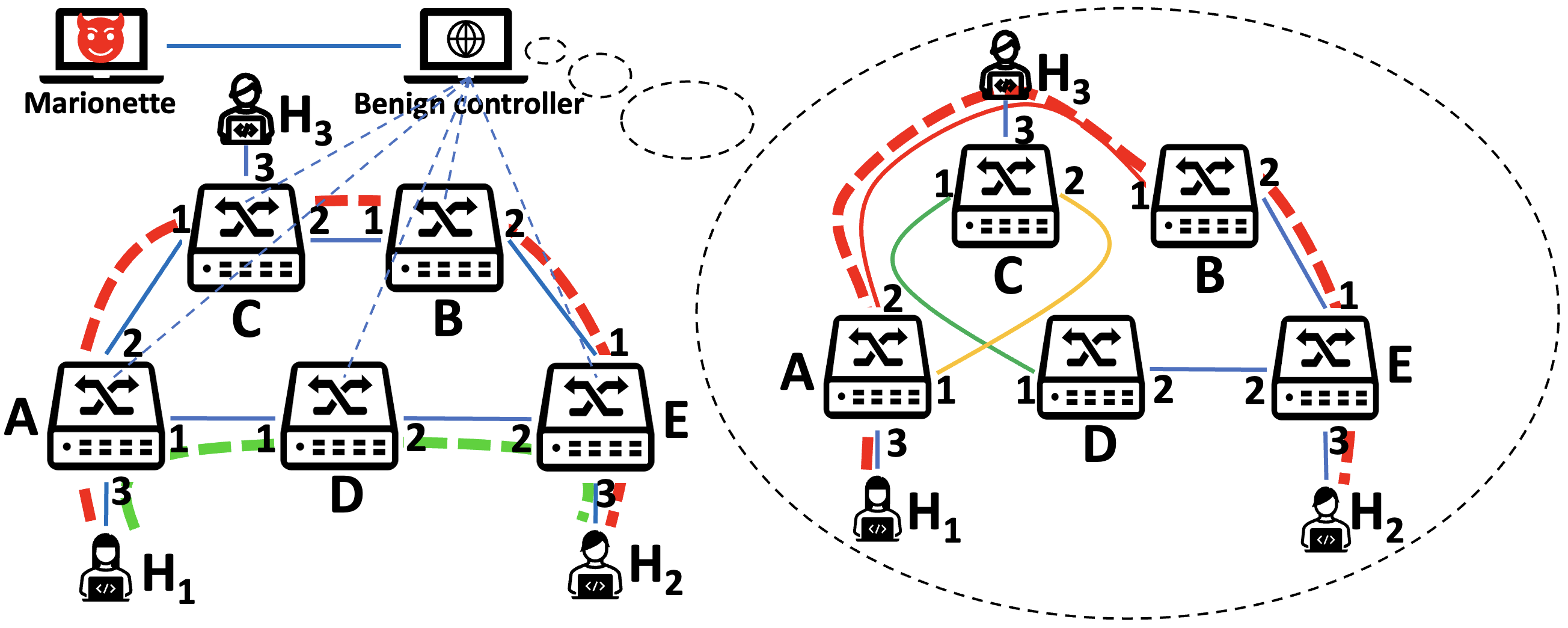}} 
    \caption{Motivating Example}    
    \label{fig:motivation_example}
    \vspace{-10pt}
\end{figure}

\inlinedsection{\update{Motivating Example}} Figure~\ref{fig:motivation_example} exemplifies this attack. 
The left-hand side shows the \textit{real} topology, while the right-hand side (i.e., the dotted circle) is the \textit{deceptive topology} as it \textit{appears} to the benign controller. Before the attack, the flow $f_1:(src:H_1,dst:H_2)$ was routed through the shortest path of green dotted path $H_1\rightarrow A\rightarrow D\rightarrow E \rightarrow H_2$  on the real topology. 
The deceptive link of $A\rightarrow B$ on the deceptive topology in Figure \ref{fig:motivation_example} can be fabricated by inserting a flow entry at switch $C$ that instructs $LLDP_{A2}$ to be forwarded to Port 2 instead of feeding it back to the controller. Consequently, the benign controller thinks the shortest path of $f_1$ is the red dotted path $H_1\rightarrow A\rightarrow B\rightarrow E \rightarrow H_2$ based on the deceptive topology. The absence of $C$ on the fabricated link of $A\rightarrow B$ allows $C$ to eavesdrop on $f_1$ without being noticed by the benign controller (red dotted path on the real topology). The details of precise link manipulation are in \S\ref{sec:approach}.
Note that the deceptive topology in this work always has the same degree sequence as the real topology to remain stealthy. We show an implementation of this attack against the ONOS controller in \S\ref{sec:eval_controller} and \S\ref{sec:eval_cluster}.

\subsection{OpenFlow and OpenFlow Link Discovery}

\update{OpenFlow~\cite{openflow} is a standard SDN protocol that allows a controller to gather information and instruct switches on how to route incoming packets by sending them flow entries\footnote{The flow entry has a \emph{match} field and a set of \emph{actions} to apply to the matched packets. Entry types for matching include source/destination MAC and IP addresses, VLAN ID, input port, etc. Instructions associated with each entry may contain actions instructing packet forwarding, packet modification, group table processing, etc.}. When a packet arrives at a switch, the switch checks its flow table to determine whether there is a matching flow entry to determine where to forward the packet. If yes, the switch forwards the packet accordingly. Otherwise, it encapsulates the packet in a \kw{packet-in} message and sends it to the controller due to a table-miss flow entry. A table-miss flow entry is the default flow entry instructing the unmatched packet to be sent to the controller. Once the controller determines the route, it sends the packet back to the switch encapsulated in a \kw{packet-out} message and installs necessary flow entry\footnote{When the controller fails to determine an output at the current switch (because the destination host is unknown or the switch is not on the computed route), the controller sends the packet back to the switch and instructs it to flood this packet through all its ports.} in the switch using \kw{flow-mod} messages. The controller repeats this process until the packet arrives at its destination. This process is called \emph{reactive forwarding}. The controller can also configure flow entries beforehand spontaneously, which is called \emph{proactive forwarding}.}

\begin{figure}[t]
  \centering
  \begin{minipage}[b]{0.23\textwidth}
    \includegraphics[width=\textwidth]{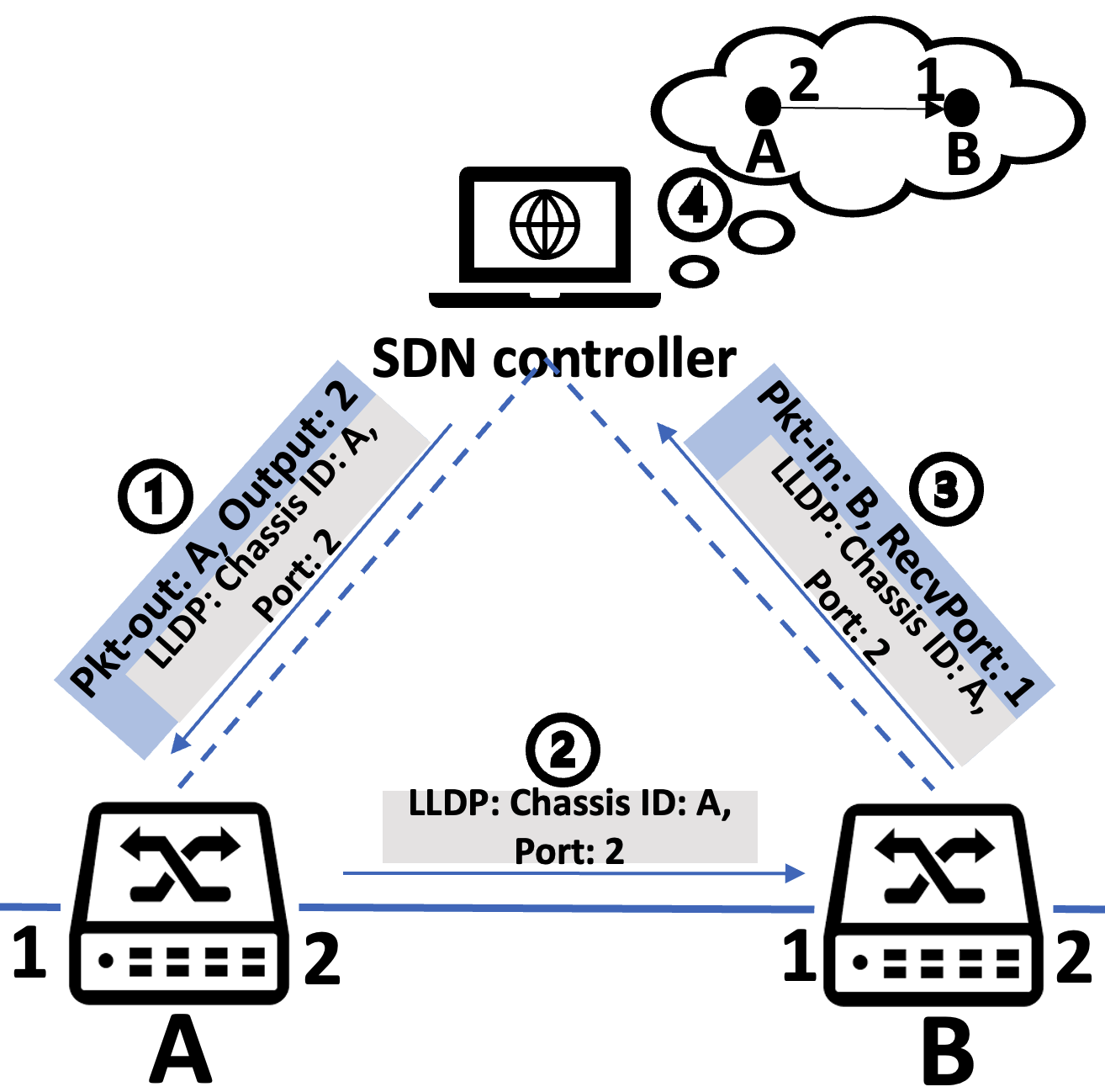}    
    \caption{OFDP illustration}
    \label{fig:ofdp}
  \end{minipage}
  \hfill
  \begin{minipage}[b]{0.22\textwidth}
    \includegraphics[width=\textwidth]{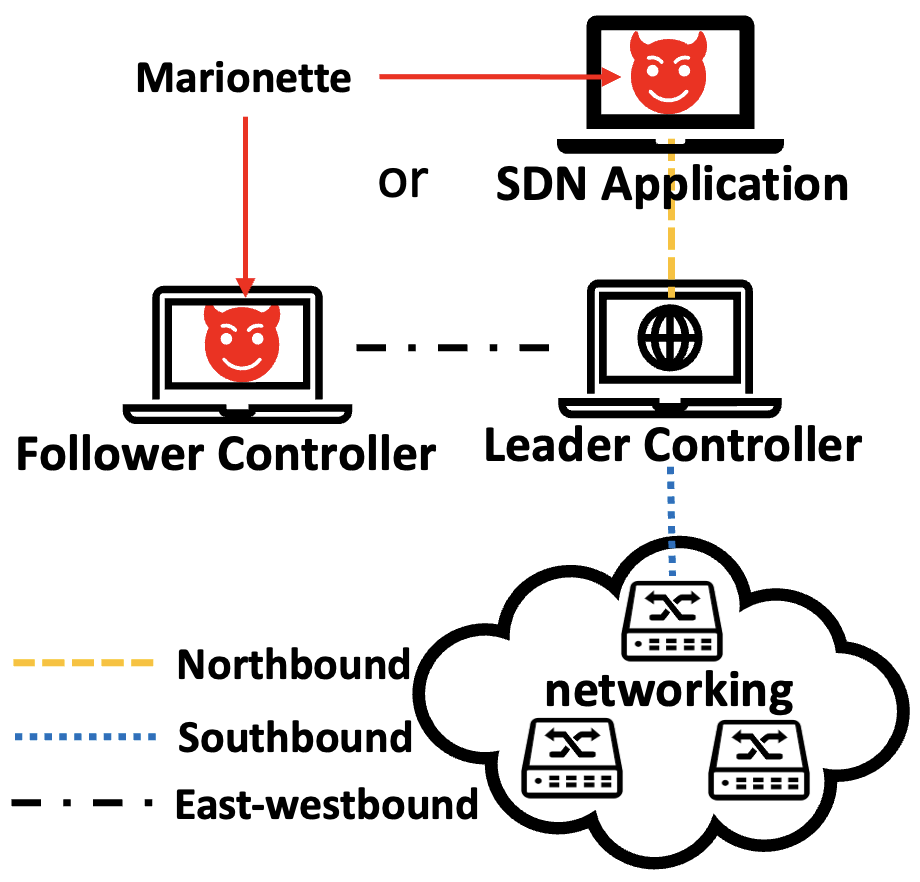}    
    \caption{Threat Model}
    \label{fig:threat_model}
  \end{minipage} 
  \vspace{-10pt}
\end{figure}

To maintain correct centralized control of modern, dynamically changing networks, controllers (re-)discover the network topology frequently (e.g., commonly every 100 ms), via the OpenFlow Discovery Protocol (OFDP).
In Figure~\ref{fig:ofdp}, we illustrate a single instance of OFDP discovering a unidirectional link from \switch{A} to \switch{B}. The OFDP process is initiated in step \ding{192} when the controller sends a \kw{packet-out} message encapsulating an LLDP packet, which we denoted as $LLDP_{A2}$, to switch~\switch{A} and instructs \switch{A} to forward the LLDP to port 2. 
In step \ding{193}, switch~\switch{A} receives the \kw{packet-out} message, decapsulates the OpenFlow header, and sends the $LLDP_{A2}$ packet to port 2; switch~\switch{B} receives the $LLDP_{A2}$ packet sent from (\switch{A}, Port: 2) (\ding{194}).
The LLDP packet matches the pre-configured table-miss /LLDP flow entry\footnote{match: ``ether-type:0x88cc", where ``0x88cc" is LLDP packet type} in switch~\switch{B} which instructs the switch to send it to the controller using a \kw{packet-in} message. Finally, in step \ding{195}, the controller receives the \kw{packet-in} message. It learns that the $LLDP_{A2}$ packet was sent from (\switch{A}, Port:2) by checking the $LLDP_{A2}$ payload, and received by (\switch{B}, Port:1) 
by checking the \kw{packet-in} payload. The controller writes a unidirectional link (\switch{A}, Port:2) $\rightarrow$ (\switch{B}, Port:1) to its data store. Hereafter, we use the notation $A2\rightarrow1B$ for simplicity. 

There have been many modifications to OFDP to improve performance. Table~\ref{table_protocols} (\S\ref{sec:eval_protocol}) lists all wire SDN discovery protocols~\cite{wazirali2021sdn} and analyzes our attack against each protocol.

\subsection{\textsc{Marionette}}\label{sec:marionette}

\projname{} is a new topology poisoning attack that uses poisonous OpenFlow~\cite{openflow} flow entries to manipulate link discovery packet forwarding to induce benign controllers to discover a poisoned topology independently.
Our key insight is that SDN discovery protocols such as the OpenFlow Discovery Protocol (OFDP) rely on a controller flooding the network with discovery packets to infer the topology. These packets are fed back to the controller from switches by either a table-miss or an LLDP flow entry.  The controller knows the start and end points of a discovery packet's traversal; however, it cannot discern any intermediate points along the feedback loop, because that information is not stored in the LLDP packets. This creates an opportunity for a rogue controller to stealthily manipulate the path taken by the discovery packets which in turn manipulates the links that are discovered by the controller.

The attack requires two types of flow entries to be installed in a switch.  The first type, which we call \textit{poisonous flow entries}, is used to misdirect topology discovery packets so that legitimate controllers learn the poisoned topology independently.  The second type, which we call \textit{gap-patching flow entries}, is used to repair the updated paths (based on the poisoned topology) that are disjoint in reality but are seen as connected by the legitimate controllers, so that packets on these paths can still be routed end-to-end.

\begin{table*}[hbt!]
\caption{Related Threats and State-of-the-art Defences}\label{table:related_threat}
\footnotesize
\begin{tabular}{p{1.4cm}p{3.8cm}p{1.28cm}p{4.7cm}|p{2.4cm}|p{2cm}}
    \toprule
    \textbf{Attack Surface} & \textbf{Attack Methodology}  &
    \textbf{Impact} & \multicolumn{3}{p{6.8cm}}{\textbf{Defense}} \\
    \midrule
     Malicious Host(s)/ Switch     
     & \multirow{5}{3.9cm}{\attack{1} 
     \textbf{Link Fabrication:}
     Fabricate a deceptive link A1 -> B2 on the controller's view. \\
     
     Methods: 1. Relay LLDP packet from A1(switch A port 1) to B2 by: 
     \begin{enumerate}[leftmargin=1.4em]
         \item malicious hosts~\cite{hong2015poisoning,skowyra2018effective}
         \item malicious switch~\cite{alimohammadifar2018stealthy}  \\
         ("Match: ether-type: 0x88cc, inport:1 Action: 2"~\cite{al2022link})
     \end{enumerate}
     2. Fabricate LLDP packet with "src: A1" and send it to B2 by:
     \begin{enumerate}[leftmargin=1.4em]
        \setcounter{enumi}{2}
         \item malicious hosts~\cite{hong2015poisoning,skowyra2018effective}
         \item malicious switch~\cite{baidya2020link}
     \end{enumerate}}
      & Localized Persistent
      & \multicolumn{3}{p{9.8cm}}{\defense{1} SPV: With benign controller assumption, it sends a probe to go through the target link. If the link is fabricated, the probe forwarding through A1 will be dropped or fed back to the controller by a switch port other than B2~\cite{alimohammadifar2018stealthy}.}\\
      \cmidrule{4-6}
      & &  & \multicolumn{3}{p{9.8cm}}{\defense{2} Latency-based Detection: A link is considered fabricated if its latency is greater than a threshold value~\cite{smyth2017detecting,huang2020towards,al2022link,gao2022defense}.}\\
      \cmidrule{4-6}
      & &  & \multicolumn{3}{p{9.8cm}}{\defense{3} Sphinx: With benign controller assumption, it builds the flow graph incrementally from Flow-mod OF-message, comparing it with actual flow routings using switch metadata. If a discrepancy is found, fabricating links are on the differing points~\cite{dhawan2015sphinx}.}\\
      \cmidrule{4-6}
      & &  & \multicolumn{3}{p{9.8cm}}{\defense{4} Port-based Detection: If an active port connects more than one link, either switch-host link or switch-switch link, one of the switch-switch links is fabricated~\cite{baidya2020link,dhawan2015sphinx}.}
     \\
     \cmidrule{4-6}
     & &  & \multicolumn{3}{p{9.8cm}}{\defense{5} TopoGuard(+): If an active port connects with both host and switch, the link connects switch is fabricated; Authenticating LLDP packets to prevent LLDP packet fabrication~\cite{hong2015poisoning,skowyra2018effective}.} \\
    \cmidrule{1-6}
    
    \multirow{3}{1.6cm}{Malicious Application/ Controller in Cluster} &  \multirow{2}{3.8cm}{\attack{2} \textbf{Wrong Information Sharing:} Malicious controller shares wrong (topology) information in the controller cluster to cause topology poisoning~\cite{khan2016topology, ropke2018preventing, maleh2023comprehensive}.} & Globalized Ephemeral & \defense{6} Dynamic Cluster: Leader controller is dynamically assigned~\cite{hu2018multi,zhang2018survey,tootoonchian2010hyperflow} and it periodically (e.g., every 100 milliseconds ~\cite{wazirali2021sdn})
    re-discovers the topology from the network independently~\cite{wazirali2021sdn,azzouni2017softdp,dhawan2015sphinx}. & \multirow{2}{2.5cm}{\defense{7} Monitoring-based Detection: Defenses~\cite{zhou2018sdn,ujcich2018cross,ujcich2021causal,ropke2018preventing} audit the controller's or applications' behavior to detect anomalies by backup controller, information flow model, root causes, or comparison with real network state.}& \multirow{2}{2.2cm}{\defense{8} Voting-based Defense: Fleet~\cite{matsumoto2014fleet},  \textsc{topotrust}~\cite{adjou2022topotrust}, and Mcad-SA\cite{qi2016intensive} depend on the majority to vote out a malicious controller based on odd behaviors around reactive forwarding or spoofing. }\\
    
    \cmidrule{2-4}
    & \multirow{3}{3.8cm}{\attack{3} \textbf{Malicious Routing:} Insert malicious flow entries to conflict with existing traffic-routing flow entries to detriment flow routings, network policy, or other applications~\cite{khurshid2012veriflow, porras2012security, al2010flowchecker,kazemian2013real, ujcich2018cross, habib2022mitigating}.}& Globalized Persistent & \defense{9} Flow Rule Checker: FlowChecker~\cite{al2010flowchecker}, Veriflow~\cite{khurshid2012veriflow},  Netplumber~\cite{kazemian2013real}, Eirene~\cite{habib2022mitigating}, and FortNOX~\cite{porras2012security} build the flow graph/logic with header information (e.g.src/dst IPs/ports) of flow entries to detect flow rule conflicts and policy violations. & &\\

    \cmidrule{2-6} 
    & \textbf{\projnameTable{}}: Insert poisonous flow entries to bypass table-miss/LLDP flow entries to poison the topology. & Globalized Persistent & \multicolumn{3}{p{9.8cm}}{\textbf{No existing defense against \projnameTable{}} because: existing defenses have overlooked the vulnerability where regular flow entries conflict with table-miss and LLDP flow entries (which is called priority-bypassing attack more specifically) to manipulate link discovery results, as detailed in Table \ref{table:features}.
    }\\
    \bottomrule
\end{tabular}
\end{table*}

To illustrate, using compact notation, given a segment $A2-1C2-1B$ in the real topology in our motivating example, our goal is to make $(A, Port:2)$ appear to connect to $(B, Port:1)$ directly as $A2-1B$. Note that since we are removing a link from the topology to connect $A$ directly to $B$, we need to add additional links back to the idle ports to go undetected. As a result, the links $A1-2C$ and $C1-1D$ are also fabricated. The example assumes LLDP packets are used in the topology discovery protocol, but our attack works on most of the existing OpenFlow-based discovery packets listed in Table \ref{table_protocols} (\S\ref{sec:eval_protocol}) without any required changes.

\section{Related Work}\label{sec:related_work}

\update{Table~\ref{table:related_threat} summarizes previously developed topology poisoning attacks and the defenses that have been proposed to defend against them. Table \ref{table:features} highlights the features in \projname{} that allow it to evade these defenses. We refer to the table entries in our discussion below.  Prior attacks can be categorized as those that fabricate links (A.1), share false topologies (A.2), and maliciously route packets (A.3). Our attack is unique in that it exploits a latent vulnerability that has been overlooked by prior attacks and defenses: \emph{flow entries intended for traffic forwarding can impact link discovery results}.
}

\inlinedsection{Link Fabrication Attacks (A.1)}
The first set of attacks originate from malicious hosts or switches and are typically localized in their network impact.  Research in this area focuses on link fabrication attacks~\cite{hong2015poisoning}\cite{skowyra2018effective}\cite{alimohammadifar2018stealthy} and includes two key techniques: (1) Fabricating LLDP packets; (2) Relaying LLDP packets. 
Sungmin, et al.~\cite{hong2015poisoning} and Richard, et al.~\cite{skowyra2018effective} show that a malicious host connecting with a switch can fabricate a link by analyzing the received discovery packets, fabricating a discovery packet, and sending it back to the switch. Another approach to link fabrication is relaying legitimate LLDP packets to a wrong endpoint using malicious hosts~\cite{skowyra2018effective} or switches~\cite{alimohammadifar2018stealthy}. 
\projname{} is different from existing works because it is initiated from the control plane and does not need to spoof or fabricate packets.   By contrast, \projname{} does not fabricate or relay LLDP packets, but uses poisonous flow entries to deceive controllers in a cluster into accepting a new poisoned topology.  Our attack is global, persistent, and stealthy. 

\inlinedsection{Link Fabrication Defenses (D.1-5)}
\textsc{TopoGuard}~\cite{hong2015poisoning} and \textsc{TopoGuard+ (D.5)}~\cite{skowyra2018effective} record and verify the identity of connected switches to mitigate host-involved link fabrication attacks. These methods do not detect \projname{} because it is not launched from the edge of the network. 
Defenses that use port-based detection (D.4)~\cite{dhawan2015sphinx,baidya2020link} verify that active ports are connected to only one other active port. We maintain this property by designing the poisonous topology with the same degree sequence as the real topology.
Latency-based detection methods (D.2) ~\cite{smyth2017detecting,huang2020towards} analyze link latencies and consider a link fabricated if its latency is greater than a threshold value.
\update{This method suffers from a high error rate due to link latency fluctuation.  \projname{} only uses switches that have undiscernible incremental latency, especially when the network latency is high~\cite{gao2022defense,al2022link}, to relay LLDP packets. }

\textsc{Sphinx (D.3)}~\cite{dhawan2015sphinx} assumes controllers are benign and trusts them to construct a flow graph database to detect abnormal flows.
Similarly, stealthy probe-based verification \textsc{SPV (D.1)}~\cite{alimohammadifar2018stealthy} trusts the control plane and uses it to probe target links.  As a result, it cannot detect our attack as \projname{} is initiated from the control plane, which both solutions trust, and is able to access the probe information to proactively patch gaps.

\inlinedsection{Malicious Controller/Application Attacks (A.2-3)} The second set of attacks in the literature assumes a malicious controller in a cluster, or a rogue SDN application.  As a result, these attacks affect the network globally.  Some malicious controllers will share incorrect topology information~\cite{khan2016topology, ropke2018preventing, maleh2023comprehensive} with the other controllers in the cluster. While these attacks wreak havoc, they do not last long because controllers periodically re-discover the topology such that the wrong topology is replaced with the correct one.  Another class of attacks inserts malicious flow entries that impact the routing of network traffic~\cite{khurshid2012veriflow, porras2012security, al2010flowchecker,kazemian2013real, ujcich2018cross, habib2022mitigating}.  Such attacks are persistent; however, there are many systems available to detect them.

\begin{table*}[t]
\caption{\projnameTable{} Features and Evasion of Existing Detection Mechanisms}
\footnotesize
\label{table:features}
\renewcommand{\arraystretch}{1.2}
\centering
\begin{tabular}{p{4.1cm}p{3.7cm}p{9cm}}
    \toprule
    \textbf{Feature} & \textbf{Related Defense} & \textbf{Evading Detection Mechanism}\\
    \midrule
    Fabricates links without spoofing, hiding, or fabricating packets &\defense{5} TopoGuard(+), \defense{7} Monitoring-based Detection & \defense{5} detects fabricated packets, and \defense{7} identifies spoofing and hiding behaviors. \textbf{\projnameTable{} does not engage in any such activities. }\\
    \midrule
    Explore the vulnerability that flow entries can manipulate link discovery precisely &\defense{7} Monitoring-based Detection & \defense{7} analyzes control plane events  to identify suspicious behaviors but overlooks \textbf{the vulnerability explored by \projnameTable{} that flow entries meant for traffic routing can manipulate link discovery.}\\
    \midrule
    Originates at malicious controller/application  & \defense{1} SPV, \defense{3} Sphinx& \textbf{\projnameTable{} is initiated from control plane} that is trusted by \defense{1}\defense{3} to detect anomalies initiated from data plane.\\
    \midrule
    Utilizes flow entries matching MAC address (with in-port) with higher priority to bypass table-miss/LLDP flow entries &\defense{9} Flow Rule Checker  &\defense{9} examines flow entry conflict with existing flow entries that routes data plane traffic.  
    However, 
    \update{\defense{9} fails to detect \projnameTable{} because \textbf{the poisonous flow entries by \projnameTable{} conflict with table-miss/LLDP flow entries, which are not monitored by \defense{9}.}}
    \\
    \midrule
    Proactively poisons topology with stealthy flow entries & \defense{8} Voting-based Defense & \textbf{\projnameTable{} leaves benign controllers in \defense{8} to incorrectly process reactive forwarding independently} due to altered topology.\\
    \bottomrule

    Instructs switches to relay link discovery packets  & \defense{2} Latency-based Detection & \update{\defense{2} 
    has a high rate of misjudgment, which may fail to detect the \textbf{low-latency switch-based relay of \projnameTable{}} very often~\cite{al2022link,gao2022defense}.}\\
    \hline
    
    Maintains the same degree sequence and high graph similarity &\defense{4} Port-based Detection, \defense{5} TopoGuard(+) & \textbf{Preserved degree sequence by \projnameTable{}} ensures each active port connects to only one link, making \defense{4}\defense{5} ineffective in detection.\\
    
    \hline
    Any controller in cluster discovers deceptive topology independently & \defense{6} Dynamic Cluster & \textbf{\projnameTable{} allows any dynamically assigned leader controllers in \defense{6} to consistently discover the same deceptive topology.} \\
    \hline
\end{tabular}
\end{table*}

\inlinedsection{Malicious Controller/Application Defenses (D.6-9)} Monitoring-based detections~\cite{zhou2018sdn,ropke2018preventing,Gwardar18} rely on tracking the controller/ application behaviors by the backup controller or intercepted OpenFlow messages and comparing them with real network state to detect anomalies. SDN-RDCD (D.7)~\cite{zhou2018sdn} detects real-time tampering-based attacks from malicious network elements (switches and controllers) by utilizing a backup controller to audit network events to detect inconsistency among them. TopoTrust (D.8)~\cite{adjou2022topotrust} detects any spoofing and tampering-based attacks with a blockchain technique. \projname{} does not cause the type of inconsistencies these systems detect because it does not use tampering, spoofing, or hiding. 

More recently, ProvSDN~\cite{ujcich2018cross} tracks the information flow on the controller to capture cross-app attacks that cause unwanted flow entries to be installed by other applications. PicoSDN enhances ProvSDN by adding a data plane model to achieve fine-grained analysis. Nevertheless, these systems overlook the vulnerability explored by \projname{}:  flow entries meant for traffic routing can also influence topology discovery, thus failing to establish a causal connection and leaving \projname{} undetected.

VeriFlow~\cite{khurshid2012veriflow}, FlowChecker~\cite{al2010flowchecker}, and FortNOX~\cite{porras2012security}~\defense{9} depend on exploring the source/destination address or port information of existing traffic-routing flow entries 
to construct forwarding graphs to detect flow entry conflicts. However, the table-miss flow entry (with no match) and the LLDP flow entry (matching protocol: $0x88cc$) are not traffic-routing flow entries, and are thus not monitored by them. As a result, \projname{} poisonous flow entries will not be detected as a conflict because they do not conflict with the monitored traffic-routing flow entries but the unmonitored table-miss and LLDP flow entries. 
Similarly, NetPlumber~\cite{kazemian2013real} employs header space analysis to construct a plumbing graph capturing all possible flow paths to ensure incoming flow entries comply with predefined network flow policies. \projname{} uses poisonous flow entries to impact link discovery results and has no signature yet.

Unlike existing malicious routing attacks \attack{3} which insert malicious flow entries to alter data plane traffic routing directly~\cite{khurshid2012veriflow, al2010flowchecker,kazemian2013real, ujcich2018cross}~\footnote{Because each flow entry has a priority value such that the highest priority match takes precedence when a packet may match multiple flow entries, a malicious flow entry with higher priority can conflict/override existing flow entries to manipulate flow routing.}, the flow entries inserted by \projname{} match \kw{ether-src} (with \kw{in-port}) to supersede table-miss/LLDP flow entries thus manipulating LLDP packet forwarding. This results in the benign controller independently routing traffic incorrectly due to the false topology. 
This attack is understudied and is undetectable by existing defenses. 

Table~\ref{table:features} summarizes the features of \projname{} and justifies why these features enable \projname{} to evade detection by existing defenses.

\section{Threat Model}
The goal of this work is to persistently poison the topology view of SDN controllers to alter some routes to enable traffic eavesdropping or flow exclusion from monitoring at a switch. The control plane topology poisoning attack is an indirect data plane attack that can be initiated from either a malicious controller in a fault-tolerant multi-controller scenario or a malicious SDN application above the control plane (see Figure \ref{fig:threat_model}). Unlike existing link fabrication attacks~\cite{hong2015poisoning,skowyra2018effective,alimohammadifar2018stealthy}, our approach does not require any switches or hosts to be compromised or malicious. The victims are benign controllers and the rest of the impacted network.

\inlinedsection{Application-Controller Scenario} 
In this scenario, a malicious application runs on top of a secure control plane---an attack vector popular in existing works~\cite{cao2020match,ujcich2018cross,ujcich2021causal,kim2023intender}.
SDN is designed to support third-party applications. These applications may be a REST client~\cite{kim2023intender} using the northbound interface\footnote{The Restful northbound API is supported by both OpenDaylight and ONOS controller to support diverse applications even written in diverse languages.} or originate from third-party developers\footnote{The open-source SDN controller projects (e.g. OpenDaylight and ONOS) allow third-party developers to submit their applications to be included in the official repositories. What's worse, some commercial products (e.g. Samsung SDN solution~\cite{samsung}, Cisco ACI~\cite{cisco_aci}, Comcast~\cite{ONOS_telecom}) are developed based on open-source controllers. Network solution companies also provide applications to run on their SDN controllers (e.g. Junos Space~\cite{junos}, Cisco ACI~\cite{cisco_aci}).} and are thus untrusted and potentially malicious~\cite{ujcich2018cross}.  Attackers may also phish to repackage or redistribute malicious installers that infect controllers at runtime~\cite{blackhat}.
Configuring flow entries is fundamental for many SDN applications (e.g. routing applications). Both OpenDaylight and ONOS provide northbound RESTful APIs for configuring flow entries.
Attacks originating from SDN applications necessitate basic privileges, including read permissions to topology, nodes (switches), and flow table information, and write permission to the flow table. Notably, the malicious SDN application does not exploit northbound API vulnerabilities or attempt to exhaust controller resources. Instead, it utilizes these privileges normally to evade existing malicious SDN application defense strategies~\cite{lee2018indago,ropke2018preventing}.

\inlinedsection{Multi-Controller Scenario} This scenario comprises a fault-tolerant multi-controller SDN architecture in which one of the controllers is malicious. \update{Attackers have leveraged rogue package managers or controller source code repositories~\cite{blackhat} to compromise a controller, and controller impersonation attacks~\cite{maleh2023comprehensive, kreutz2014software} (e.g., Appendix \ref{sec:eval_impersonation}) to introduce a malicious controller in the cluster. 
The minimum required privileges of the malicious controller include \emph{read} permissions to the topology, nodes (switches), and flow table data store, along with \emph{write} permissions to its own flow table data store.} 
\update{
These requirements align with the fundamental capabilities of OpenDaylight's and ONOS's fault-tolerant replication implementation and can be fulfilled without necessitating the leader role within the fault-tolerant cluster. This is because a malicious controller can induce changes in the network topology via (1)~the direct manipulation of flow entries in the OpenFlow switches when it has an equal role~\footnote{
OpenFlow v1.3~\cite{openflow} defines three roles for a controller: leader, follower, and equal. A follower controller has read-only access to the switch and does not receive asynchronous messages(e.g. \kw{packet-in}). Both leader and equal controllers can modify the switch state and receive asynchronous messages from the switch. }, or through (2)~a confused deputy attack when it has a follower role, in which flow entries are altered in the malicious controller's datastore, causing a leader controller to learn and deploy these entries through consensus.
Consequently, the malicious controller can operate with any (leader, equal, or follower/none) role to launch the \projname{} attack.}

\section{\textsc{Marionette} Flow Entry Design}\label{sec:approach}
This section details \projname{}'s flow entry design using the example presented in \S\ref{sec:overview_motivation}.  \trent{Update this sentence too.}

\subsection{Precise Link Manipulation}\label{sec:simple_pois_entry}
The flow entries that instruct data plane traffic forwarding can also be used for discovery packet forwarding.  Existing discovery protocols typically utilize the default table-miss flow entry or a flow entry matching the LLDP packet type (which we call LLDP flow entry) to cause incoming LLDP packets to be forwarded to the controller in a \textit{packet-in} message. Similarly, it is possible to use flow entries to manipulate discovery packet forwarding to fabricate links.  To do this, poisonous flow entries must be inserted into the switch at a higher priority than the table-miss/LLDP flow entry. 
The addition of poisonous flow entries at a higher priority than existing entries will not affect the hit rate metrics of existing entries as discussed in \S\ref{sec:vlanentries}.

Link fabrication requires the following:
\begin{requirement}\label{rule_link}
 {Given port $x$ of switch $A$ and port $y$ of switch $B$ \update{that are not directly connected}, to fabricate a link $Ax\rightarrow yB$, an attacker must force a discovery packet sourced from ($A$, Port:$x$) to be received at ($B$, Port:$y$).}
\end{requirement}

To execute the attack, we take advantage of the fact that LLDP packets do not encode intermediate link information. For this reason, the controller only knows the node and port to which it sends the discovery packet and the node and port from which the discovery packet is returned. It uses these two pieces of information from all of the discovery packets to map the network. The controller expects that each switch will send the discovery packet back to it at each hop allowing it to learn every link.  To make switch $C$ appear invisible on link $A2\rightarrow1B$ to the controller in our example, we must therefore force switch $C$ to forward the LLDP packet to $B$, rather than pass it back to the controller.  We can do so by creating a poisonous flow entry with a higher priority than the table-miss/LLDP flow entry in Switch $C$.  
For example, it is easy to construct a na\"ive poisonous flow entry in $C$ that forwards all packets of type LLDP (i.e., ethernet-type:0x88cc) to switch $B$. So, we can simply insert flow entry $e_1^C$ on $C$:
\begin{center}
\footnotesize
\begin{tabular}{|c|ll |ll@{\hspace{.2cm}} |} 
\hline
 $e_1^C$ &\textbf{match}: & ether-type: 0x88cc & \textbf{action}: & output: 2  \\ [0.5ex] 
\hline
\end{tabular}
\end{center}

However, this flow entry is easily detectable because ``0x88cc" can be recognized as LLDP and this flow entry's output is atypical for an LLDP-related flow entry.  A normal LLDP flow entry has the actions as the \kw{to\_controller} rather than a \kw{out\_port\_num}. Moreover, it is problematic because links are bidirectional ($A2\leftrightarrow1C2\leftrightarrow1B$). The poisonous flow entry $e_1^C$ on $C$ influences all LLDP packets received by $C$, so the entry can fabricate links but fails to precisely affect the bidirectional link. This may result in one port being part of two links which can be detected by existing port-based detection~\cite{dhawan2015sphinx,baidya2020link}.

To manipulate links through a stealthy set of poisonous flow entries, we rely on the MAC addresses of the source switch port to fabricate links shown as flow entries $e_2^C$ and $e_3^C$ on $C$. These entries achieve our goal without conflicting with other flow entries because all MAC addresses are unique, and only discovery packets (of all types) use the switch port MAC address as the source MAC in their header. 
\begin{center}
\footnotesize
\begin{tabular}{|c|ll|ll@{\hspace{.2cm}}|} 
\hline
 $e_2^C$ &\textbf{match}: &ether-src: A2\_mac  & \textbf{action}: & output: 2  \\ [0.2ex] 
\hline
 $e_3^C$ & \textbf{match}: &ether-src: B1\_mac  & \textbf{action}: & output: 1  \\ [0.2ex] 
\hline
\end{tabular}
\end{center}

\inlinedsection{Stealthiness}
Note that $e_2^C$ and $e_3^C$ are stealthy because nothing easily detectable is contained in the ``match" field of the flow entry. The MAC address of a switch port is nothing but a string of 12 hexadecimal digits as any host. We call the design of $e_2^C$ and $e_3^C$ a \textit{vanilla poisonous flow entry design}, and it directly attacks most of the existing SDN discovery protocols because the standard ethernet header is mandatory on any discovery packet. 

\subsection{VLAN-based Poisonous Flow Entries}\label{sec:vlanentries}
\update{However, ONOS controller uses the \kw{ether-src} field of LLDP packets to store a unified fixed fingerprint (\kw{FFP}) to identify clusters~\cite{onos_disc}.  
This limits the attack effectiveness when matching the \kw{ether-src} to distinguish LLDP packets from different sources}; therefore, we need another approach.

To realize the attack, 
we first configure $e_4^C$ and $e_5^C$ matching \kw{ether-src} with \kw{in-port} on $C$ to distinguish $LLDP_{A2}$ and $LLDP_{B1}$ to fabricate links $A2 \rightarrow 1B$ and $A2 \leftarrow 1B$:

\begin{center}
\footnotesize
\begin{tabular}{|c|ll|ll@{\hspace{.2cm}}|} 
\hline
 $e_4^C$&\textbf{match}:&{ether-src: \update{\kw{FFP}}, in-port: 1} & \textbf{action}:&output:2  \\ [0.2ex] 
\hline
 $e_5^C$&\textbf{match}:&ether-src: \kw{FFP}, in-port: 2  & \textbf{action}:& output:1  \\ [0.2ex] 
\hline
\end{tabular}
\end{center}

However, $e_4^C$ and $e_5^C$ are problematic for later link fabrications. For example, in order to fabricate $A1 \rightarrow 2C$, we need to route the $LLDP_{A1}$ to be received at ($C$, Port:2) ($A1\rightarrow1D2\rightarrow2E1\rightarrow2B1\rightarrow2C$) based on the real topology. 
Irrespective of the configuration on $D$, $E$ and $B$, suppose that the $LLDP_{A1}$ has been forwarded to ($C$, Port:$2$); $C$ should forward it to the controller to complete link $A1\rightarrow 2C$ discovery, but it will forward $LLDP_{A1}$ to $Port:1$ due to $e_5^C$. To address this problem, we can use virtual local area network (VLAN) tagging to distinguish the sources of LLDP packets.  
Each VLAN has its own ID ($vlan\_id$), and we use this attribute to tag a set of flows passing through the switch, and then match on them.  To tag flows, we use the action \kw{push-vlan} in our flow entry as shown in $e_6^D$.  This entry is configured on switch $D$ and is used to tag the sources of $LLDP_{A1}$.
\begin{center}
\footnotesize
\begin{tabular}{|c|ll|ll@{\hspace{0.1cm}}|} 
\hline
 $e_6^D$&\textbf{match}:&ether-src:\kw{FFP}, in-port: 1 &\textbf{action}:&push-vlan:1, output:2 \\[0.2ex] 
\hline
\end{tabular}
\end{center}
By pushing \kw{vlan\_id}s to the LLDP packets at their first hop, the sources are marked precisely without any flow entry conflicting issues. The \kw{push-vlan} action is therefore also added to $e_4^C$ and $e_5^C$. 
As the last step of fabricating $A1 \rightarrow 2C$, the flow entries $e_7^E$ on $E$, $e_8^B$ on $B$ and $e_9^C$ on $C$ listed below are needed to collaborate with $e_6^D$ on $D$, 
and $e_9^C$'s priority must be higher than $e_5^C$ on $C$ to function correctly. $e_7^E$ matches on the \kw{vlan\_id} and forward to $B$ through $(Port:1)$. Similarly, $e_8^B$ matches on the \kw{vlan\_id} and forward to $C$ through $(Port:1)$. 
$e_9^C$ removes the \kw{vlan\_id} using the \kw{pop-vlan} action and sends the original $LLDP_{A1}$ to the controller as desired.
\begin{center}
\footnotesize
\begin{tabular}{|c|ll|ll@{\hspace{.2cm}}|} 
\hline
 $e_7^E$&\textbf{match}:&vlan-id: 1 & \textbf{action}: & output:1 \\[0.5ex]
\hline
\hline
 $e_8^B$&\textbf{match}:&vlan-id: 1 & \textbf{action}: & output:1 \\[0.5ex]
\hline
\hline
 $e_9^C$&\textbf{match}:&vlan-id: 1 & \textbf{action}: & pop-vlan, to\_controller \\[0.5ex]

\hline
\end{tabular}
\end{center}

We refer to these types of entries as \textit{VLAN poisonous flow entry design} ($e_6^D$ and $e_9^C$ are called $E_{VStart}$ and $E_{VEnd}$ respectively, $e_7^E$ and $e_8^B$ are called $E_{VBody}$) and define the path $1D2\rightarrow 2E1\rightarrow2B1\rightarrow 2C$ for fabricating $A1 \rightarrow 2C$ as the \textit{poisoning path}. The VLAN poisonous flow entry design is also important even when \kw{ether-src} is available to distinguish the source of the LLDP packet because the VLAN tunnel can be reused for gap-patching (described in Section \ref{sec:gappatchentries}).

To fabricate the same link,
the number of poisonous flow entries matching on \kw{ether-src} is less than the case of matching on [\kw{ether-src}, \kw{in-port}] because (1) VLAN is not mandatory for the single-hop case $A2-1B$, thus no \kw{vlan\_id} is needed and we can use $e_2^C$ and $e_3^C$ safely without flow entry conflicts, and (2)~the $LLDP_{A1}$ can be forwarded to the controller with a default table-miss flow entry instead of configuring $e_9^C$ to avoid flow entry conflicts. Thus, the \kw{vlan\_id} can be safely popped at $B$ by adding \kw{pop-vlan} to the $e_8^B$.

\begin{center}
\footnotesize
\begin{tabular}{|c|ll|ll@{\hspace{0.2cm}}|} 
\hline
 ${e_8^B}'$&\textbf{match}:&vlan-id: 1 & \textbf{action}: & pop-vlan, output:1 \\[0.5ex]
 
\hline
\end{tabular}
\end{center}

\inlinedsection{Stealthiness}
Each flow entry in a flow table has a corresponding metric, which increments each time a particular entry routes a packet.   As a result, to make deceptive flow entries as stealthy as possible, we must ensure that the new entries do not interfere with the counting rates of existing ones. Our distinguishable \kw{ether-src} VLAN approach keeps these counts consistent because the original table-miss/LLDP flow entries are always used to forward (original or relayed) LLDP packets back to the controller to maintain the same degree sequence of the topology. \update{But when we have to use [\kw{ether-src}, \kw{in-port}] to fabricate links precisely, the metric of original table-miss/LLDP flow entries cannot remain consistent because the mis-forwarded LLDP packet has to be fed back to the controller by our flow entries as ${e_9^C}$, instead of original table-miss/LLDP flow entry.}

Note that the distinguishable {\kw{ether-src} VLAN poisonous flow entry} design is the \textit{default} setting for the following discussion unless specified differently.
The route of this motivating example is simple with only one hop length. In real network scenarios, there could exist multiple switches along a route $Ax\rightarrow yB$ that need to be configured.  This poses a non-trivial challenge and is discussed in Section~\ref{sec:pois_link}.
\subsection{Gap Patching Flow Entries}\label{sec:gappatchentries}

Once the poisonous flow entries are set, the controllers 
believe that switch $A$ directly connects to switch $B$, when in reality switch $C$ is in the middle. As a result, when inserting flow entries for paths that include the deceptive link between $A$ and $B$, the controller will not configure a corresponding flow entry on $C$ to complete the path because it is unaware that $C$ is on the path. This results in a gap in the path configuration.

We fix this issue through a \textit{gap patching} process. The idea is that for each fabricated link setup, we must ensure that the first hop on the fabricated link will route packets along the poisoning path of that link. 
To avoid flow rule conflict with the legitimate flow entries, we combine the \kw{in-port} with the original match as the match of the gap-patching flow entries to prevent any other normal forwarding packets from hitting this flow entry. 
For the single-hop gap of flow $f_1:(H1, H2)$ caused by the single-hop poisoned path, this is done through a flow entry like $e_{10}^C$.  For a longer gap of flow $f_2:(H1, H3)$, which is caused by a longer poisoning path,
we can take advantage of the existing VLAN poisonous flow entries by simply adding a flow entry that pushes the corresponding VLAN tag to the $f_2$ packets on the VLAN start switch $D$ as shown in $e_{11}^D$. In either case, only one flow entry is required to patch the gap.

\begin{center}
\footnotesize
\begin{tabular}{|c|ll|ll@{\hspace{0.2cm}}|}
\hline
 $e_{10}^C$&\textbf{match}:&ether-dst: H2, in-port:1 & \textbf{action}: & output: 2 \\[0.5ex]
\hline
\hline
$e_{11}^D$&\textbf{match}:&ether-dst: H3, in-port:1 & \textbf{action}: & push-vlan: 1, output:2 \\[0.5ex]
\hline
\end{tabular}
\end{center}

\inlinedsection{Stealthiness}
Note that we could get away without gap patching if the controller has the \textit{reactive forwarding feature} enabled.  This is because the controller will send an unknown packet back to the switch with a \kw{packet-out} message and instruct the switch to flood the packet across all ports, filling in the gap.  However, we still patch gaps even in reactive forwarding because (1)~too many \kw{packet-in} message requests sent to the benign controller at a time are suspicious; and, (2)~switch flooding unnecessarily wastes network resources. 

\update{Another note is that the gap patching is undetectable by the flow rule checker (i.e., \defense{9} in Table \ref{table:features}). Because gap patching happens after the topology is poisoned, the switches having gap patching flow entries will not be on the flow paths computed based on the deceptive topology. As a result, the flow rule detection fails to correlate the gap-patching flow entries with the flows based on its logic graph.}

\section{System Design}\label{sec:design} 

\begin{figure}[t]
 \vspace{-1\baselineskip}
\centerline{\includegraphics[width=0.7\linewidth]
{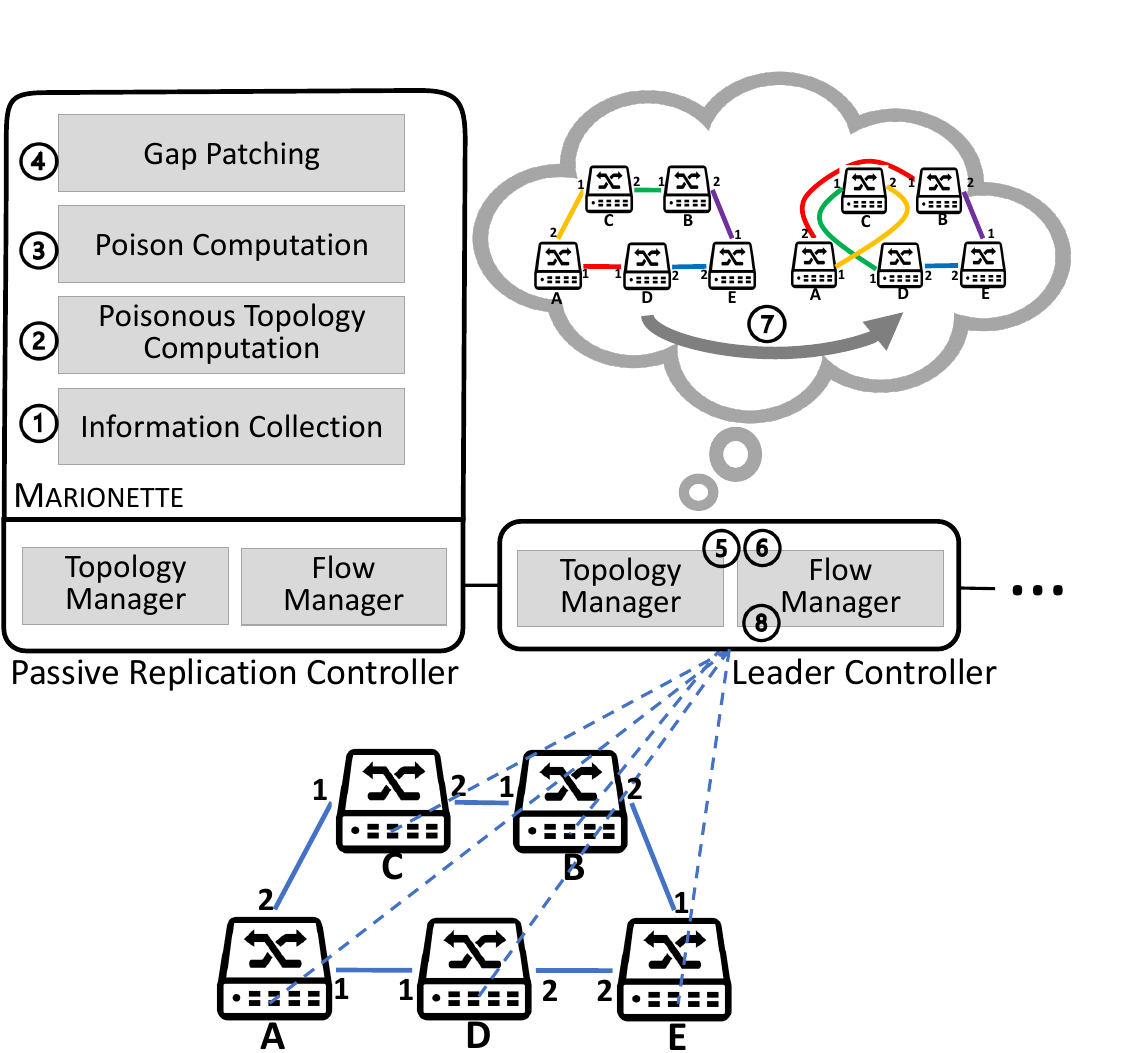}}
\caption{\projnameTable{} framework}
\label{fig:framework}
\vspace{-10pt}
\end{figure}

To automate and scale up the attack, we designed \projname{}, a framework that automatically learns an existing topology, generates a poisoned topology from an attacker's goal (e.g., evading a network monitor, conducting a man-in-the-middle attack), and generates flow entries to instantiate the poisoned topology. The framework comprises four modules: {Information Collection}, {Poisonous Topology Computation}, {Poison Computation and Setup}, and {Gap Patching}, as shown in Figure~\ref{fig:framework}. The interaction follows a sequence of eight steps.  Note that the framework assumes the presence of a malicious controller within the controller cluster, operating in the follower/none role and without direct connection to the underlying network. \update{When the \projname{} attack is initiated from a malicious application in an application-controller scenario, it follows similar steps.}

In Steps \ding{192} and \ding{193}, \projname{} learns the topology and switch information of a network from the replication controller's data store, and then uses this information as the basis for a reinforcement learning (RL) model to compute a deceptive poisonous topology. 
Step \ding{194} involves \projname{} 
composing stealthy poisonous flow entries to manipulate links based on the information gathered in Step \ding{192} to construct the computed target topology in Step \ding{193}. 
In Step \ding{195}, the Gap Patching module identifies and computes the required gap-patching flow entries for existing flows. In Step \ding{196} and Step \ding{197}, the poisonous and gap-patching flow entries are configured to the passive replication controller's data store and propagated in the controller cluster by the data store consistency mechanism. The leader controller \update{accepts these data store updates} and sends the \update{poisonous and gap-patching} flow entries to the network on behalf of \projname{}. In Step \ding{198}, the benign leader controller in the cluster independently discovers the topology which is the one designed by \projname{} from Step \ding{194}. As a result of the altered topology, the leader controller calculates and sends new routes over the deceptive topology in Step \ding{199}. The flows will arrive at their proper destinations since gap patching occurs beforehand in Step \ding{197} and the Gap Patching module keeps monitoring the flow table to patch when necessary.

In a controller cluster with an active replication design, where the malicious controller assumes an equal role, \projname{} can independently send poisonous and gap-patching flow entries while concealing them by refraining from updating the flow table data store.

\subsection{Poisonous Topology Computation}\label{sec:design_RL}

The specific attack goal determines the desired deceptive topology. We focus on two attack goals due to space considerations but can support others: (1) routing traffic to an eavesdropping switch point; (2) routing traffic away from a monitor at a given switch point.  To create a deceptive topology to satisfy these attack goals, \projname{} learns the existing topology through an information-gathering step. With the real topology and the node targeted for an attack (i.e., for eavesdropping or evasion), it employs an RL algorithm taking the collected information and the attack goal to output the desired topology.  

RL uses machine learning training methods to train an agent to accomplish some task or behavior based on trial-and-error by executing a set of actions to attain a goal.  Training is done through a feedback loop, where the RL agent is given a reward for desirable actions that lead it closer to the goal, and a penalty for actions that are counter to the goal.  The agent is effectively a state machine, and after each action, it receives a transition to the next state, and feedback from the environment in the form of a reward or penalty.

We choose RL over traditional methods because:
\begin{itemize}
    \item \textit{RL adaptation}: The reward component of the RL algorithm can be easily adapted to meet different goals, \update{without modifying the algorithm.}
    \item \textit{Flexibility}: RL provides flexibility in incorporating specific constraints relevant to the attack goals.
    \item \textit{Feasibility}: The traditional brute force exploration space can be huge: $S = {n \choose m}*({2*m \choose 2}*{2*(m-1) \choose 2}...*{4 \choose 2}-1)$ where $n$ is the total number of edges, $m$ is the number of edges allowed to be changed with a similarity constraint.
    \item \textit{Best effort}: Unlike traditional optimization formulations that may fail to find the optimal solution within a polynomial running time and return no solution, RL provides its best effort result, which is often satisfactory.
    
\end{itemize}

To create a realistic and stealthy topology using RL, we enforce two constraints: (1) maintaining the same degree sequence as the real topology, and (2) ensuring a threshold level of similarity between the poisoned and real topologies. Constraint (1) is important because distinct degree sequences are easily detected through the collection of network information via the OpenFlow protocol. Additionally, as our attack utilizes original discovery packets, \update{but does not fabricate any discovery packet,} the number of links in the poisoned topology cannot exceed the number of links in the real topology. Constraint (2) aims to prevent suspicion by avoiding sudden and significant alterations in the topology. 

\update{To accelerate the learning process, we analyze the topology characteristics and attack goal to guide the agent by certain actions~\cite{pertsch2021accelerating} which will be discussed after introducing actions of our RL model.} %

\inlinedsection{State}\label{sec:RL_state}
The state of the model is the topology after changes caused by actions. The initial state is the real topology. Mathematically, we define a port-based adjacency (PA) matrix to depict the topology:

\begin{definition}\label{def_pa_matrix}
    Given a topology with $n$ switches and $m$ bidirectional links, $p_{i,j}$ is the port number on switch $s_i$ such that switch $s_i$ connects to switch $s_j$ through port $p_{i,j}$, $n_{i,j}=(s_i,p_{i,j})$ is called the neighbor of switch $s_j$.  A port-based adjacency matrix is a $n \times n$ matrix $A$ such that:
\begin{equation}
\footnotesize
  A_{i,j}=\left\{
  \begin{array}{@{}ll@{}}
    p_{i,j}, & \text{if there is a link from switch $s_i$ to switch $s_j$}\\
    0, & \text{otherwise}
  \end{array}\right.
\end{equation} 
\end{definition}

\inlinedsection{Action}\label{sec:RL_action}
A well-known result is that graphs with the same degree sequence can convert to each other via a sequence of \textit{2-switches}~\cite{barrus20122}. 
A typical 2-switch is deleting any two edges and reconnecting the idle ports differently~\cite{jaume20202}. 
There are various ways to change the graph while maintaining the same degree sequence and they can be converted by several 2-switches. Another method to change the graph while maintaining the degree sequence is what we call \textit{node-reallocation}.
The node-reallocation is a series of 2-switches as shown in Figure \ref{fig:2_switch} which is essentially only one node-reallocation (reallocating $C$ between $A$ and $D$) but three 2-switches. Mathematically, a 2-switch is a sequence of element switching on the PA matrix shown in Figure \ref{fig:2_switch}.
The color of the matrix elements matches the color of the bidirectional fabricated links: 

\begin{figure}[t]
     \centering
     \includegraphics[width=0.45\textwidth]{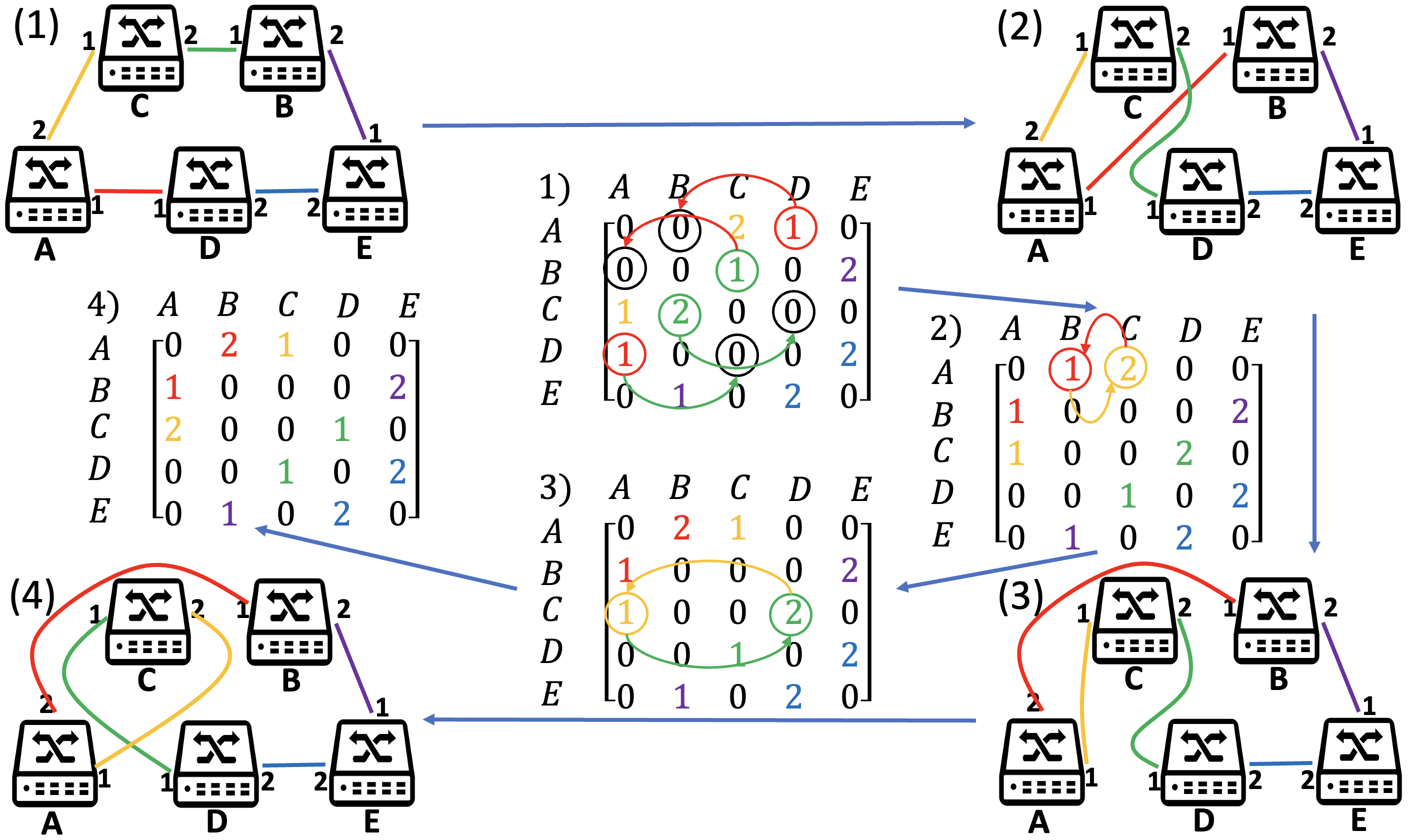}
    \caption{2-switch actions on PA matrix}
        \label{fig:2_switch}
        \vspace{-10pt}
\end{figure}

\inlinedsection[0pt]{Action Priors}\label{sec:action_prior} Both 2-switch and node-reallocation lead to any graph that has the same degree sequence. However, certain actions may work faster than others on different topologies. \update{We use this observation as the action priors~\cite{pertsch2021accelerating} to accelerate the learning process}. Intuitively, node-reallocation is well suited to force a different route to be chosen because it is the unit action that directly changes the routing distance (e.g. $A2-1B2-1E$ seems shorter than $A1-1C2-1D2-2E$ after reallocating C between $A1-1D$). However, the 2-switch action is more efficient than node-reallocation on a tree(-like) topology because there are no (less) other routes that can be forced on a tree topology due to its acyclic nature. In our evaluation,
we choose an action of either 2-switch or node-reallocation based on their suitability for different topologies. The action space size is equal to $m \choose 2$, where $m$ is the number of edges. Note that a link re-connecting to a different port on the same node is also considered as changing the adjacencies.

The number of actions $n_a$ in each training episode depends on the network size $n$ and similarity requirement $s\in[0,1]$ because an increasing number of actions in each episode results in a fabricated topology that diverges more from the real topology. We choose the number of actions roughly by $n_a \approx n\cdot(1 - s)$. 
We must also preserve the connectivity of the altered topology realistically after each action.

\inlinedsection{Reward}\label{sec:RL_reward}
We define two dimensions to measuring a reward. First, our goal is to either divert as many flows as possible to a given eavesdropping switch ($s_{vul}$) or drive as many flows as possible away from a monitoring switch point ($s_{mon}$). We refer to this maximization of flow diversions to or away from a node as our \textit{flow coverage goal}.  Second, we wish to maintain a graph similarity beyond a threshold value to be stealthy. 
For each step, when the flow coverage on $s_{vul}$ ($s_{mon}$) is greater (less) than the threshold value and the graph similarity is beyond the threshold value, 
a reward of 1 is granted. Otherwise, the reward is -1 (penalty). Episode training is complete when a step achieves the goal.
Here we prioritize the flow coverage goal and guarantee a certain graph similarity value.

\inlinedsectionit{Flow Coverage}
\update{After the topology has been altered, the flow routing will be re-calculated based on the poisoned topology. We must check whether this altered topology can trigger enough flow routing updates that meet the flow coverage goal on node $s_{vul}$/$s_{mon}$. 
Nodes such as $s_{vul}$ may not appear on the paths in the poisoned topology; however, they are still on those paths in the real topology. It is problematic to calculate flow coverage by directly checking whether the $s_{vul}$ is on the updated paths based on the poisoned topology. The correct approach is to check whether the \emph{real neighbors} of $s_{vul}$ are on the updated paths. For the example in Figure \ref{fig:motivation_example}, the updated path $P_{update}:H_1->3A2->1B2->1E3->H_2$ does not include $C$ based on the deceptive topology. But $C$ is actually between $B$ and $E$ in the real topology. We can infer this result by checking whether the real neighbors of $C$ are on $P_{update}$. $A:2$ and $B:1$ are $C$'s real neighbors and they are on $P_{update}$, so we know this deceptive topology meets our goal to eavesdrop flow $f_1$ on $C$. Finally, we count the number of such paths to get the flow coverage result. Note that the neighbor node is associated with a port number as defined in Definition \ref{def_pa_matrix}. }

\inlinedsectionit{Graph Similarity} 
There are various ways to evaluate the similarity between graphs.  We choose a simple method called vertex/edge overlap (VEO)~\cite{papadimitriou2010web} as it fits our scenario well.  VEO measures the overlapped vertices and edges between graphs. 

Since we maintain the degree sequences of the topology graph, the numbers of vertices and edges do not change.  As a result, the higher the number of vertices, the less sensitive the similarity score is to changes in the edges. To exclude this negative influence caused by the number of nodes, we derive the edge overlap (EO) method from VEO to evaluate graph similarity for our case and compute it with the following formula:
\begin{equation}\label{veo_sim}
\footnotesize
    sim_{EO}(G,G') = \frac{|E\cap E'|}{|E|}
\end{equation}
Where $G$ and $G'$ are the real topology and poisoned topology, and $E$ and $E'$ are the edge sets for both graphs. 

\inlinedsection{Stealthiness}
\projname{} modifies only a few links in a large network while maintaining the degree of each network node to evade (D.4) port-based detection. The similarity constraint is for avoiding manual detection, implementing the intuition that topologies should not change dramatically. Because some goals on certain topologies are impossible to meet while maintaining the same degree sequence regardless of the similarity, this RL model is a best-effort algorithm. The strictness of the constraints (similarity) and the effectiveness of goals (flow coverage) have a tradeoff that needs adjusting based on the situation.
\subsection{Poison Computation and Setup}\label{sec:pois_link}

Once the RL workflow creates a poisonous topology, \projname{} generates and installs the poisonous flow entries to realize the deceptive topology. In \S\ref{sec:approach}, we introduced a topology poisoning attack using poisonous flow entries and provided examples of simple link poisoning. However, in large network scenarios, the setup of poisonous flow entries can become complex due to the sheer number of possible links that can be manipulated with possible intermediate nodes along the path. To scale our approach, we designed a general algorithm for computing the necessary poisonous flow entries for each hop, taking into account the network topology, switch nodes, and a deceptive link as input. The algorithm described in Algorithm~\ref{alg:pois_link} (Appendix \ref{sec:appendix_pois_flow_entry}) implements Rule \ref{rule_link}, focusing on how to ensure that the discovery packet sent from $(A, Port:x)$ is received at $(B, Port:y)$. The details are in Appendix \ref{sec:appendix_pois_flow_entry}.

\subsection{Gap Patching Computation and Setup}\label{sec:gap_patching}

Once the topology has been poisoned, the routes calculated by the benign leader controller based on the deceptive topology may contain gaps as described in \S\ref{sec:gappatchentries}. 
We address this proactively and reactively. For proactive patching, we set the gap-patching and poisonous flow entries together because we can predict the new routes for existing flows. In reactive forwarding mode, \update{the controller instructs the switch to flood this unexpected packet~\footnote{When a packet is forwarded to the controller by a switch, but the switch's position contradicts the packet's routing path based on the recent topology, we designate it as an unexpected packet to the switch.} with \kw{packet-out} message to all its ports.} This behavior provides \projname{} with ample time to patch the gaps. In this context, the latency performance of the gap-patching model primarily affects stealthiness rather than persistence. Reducing the number of unexpected packets sent to benign controllers results in a lower level of suspicion. The details of the gap-patching algorithm are in Appendix \ref{sec:appendix_gap_patching}.

\begin{table*}[t]
\footnotesize
\caption{\projname{} attack effectiveness towards SDN Discovery Protocols}
\label{table_protocols}
\centering
\begin{tabular}[htbp]{p{1.7cm} p{10cm}p{2.2cm}p{1cm}p{1cm}}
\toprule
\textbf{Discovery Protocol} & \textbf{Methodology} &\textbf{Src MAC of Discovery PKT} & \textbf{OF-based}  & \textbf{Attack Success} \\
\midrule
OFDP~\cite{hasan2017efficient} & Controller \textbf{floods} LLDP packets to every port of every switch by Packet-out messages, build the topology based on the feedback \textit{packet-in} messages encapsulated with flooded LLDP & MAC addr of src switch port & Yes &Complete \\
\midrule
Hybrid OFDP~\cite{ochoa2015current} &  Controller \textbf{floods} LLDP and BDDP packets to every port of every switch to discover a hybrid network because Legacy switches drop LLDP packet but \textbf{broadcast} BDDP packet & MAC addr of src switch port & Yes &Complete \\
\midrule
OFDP\_v2~\cite{pakzad2016efficient} & Controller sends a single LLDP packet to every switch that is pre-configured to duplicate LLDP packets and \textbf{flood} them to all its ports &  MAC addr of src switch port & Yes &Complete \\
\midrule
ESLD~\cite{zhao2018esld} & Controller only sends LLDP packets to the switches' ports that are connected to a switch but not a host, the LLDP packets are \textbf{flooded} as in OFDP & MAC addr of src switch port & Yes & Complete \\
\midrule
BBLD~\cite{hussain2021broadcast} & Controller sends one discovery packet to the network and lets it \textbf{broadcast} in the network, tag visited ports with port status to avoid loops & MAC addr of src switch port & Yes & Complete \\
\midrule
TEDP~\cite{rojas2018tedp} & Controller sends one packet to the network and lets it \textbf{broadcast} in the shortest spanning tree, uses All-Path locking mechanism to avoid loop & MAC addr of src switch port & Yes &Complete \\
\midrule
HDDP~\cite{alvarez2020hddp} & Controller \textbf{floods} discovery packets to discover SDN nodes which is similar to OFDP, and utilizes All-path locking on the switches to discover non-SDN nodes & MAC addr of src switch port & Yes & Complete \\
\midrule
SLDP\cite{nehra2019sldp} & Controller uses random MAC source address in discovery packets to prevent packet fabrication and replay, the discovery packets are \textbf{flooded} as in OFDP & Random MAC address & Yes & Complete~\footnotemark \\
\midrule
TILAK\cite{nehra2019tilak} & Controller uses random MAC destination address in discovery packets to prevent packet fabrication and replay, the discovery packets are \textbf{flooded} as in OFDP & MAC addr of src switch port & Yes & Complete \\
\midrule
sOFTDP~\cite{azzouni2017softdp} & Controller \textbf{floods} encrypted LLDP packet to discover new links when some port is on. After that, switches use \textbf{Bidirectional Forwarding Detection(BFD)} as a port-liveness detection mechanism to quickly detect link events & MAC addr of src switch port & Partially & Restricted \\
\midrule
ForCES~\cite{tarnaras2015sdn} &  The SDN-Switches \textbf{transmit and receive} LLDP advertisements and build their topology tables without the intervention of the SDN-Controller. The controller pulls topology information from the switches & Not applicable &  No & Zero \\
\midrule
eTDP~\cite{ochoa2019etdp} & 
The switches discover the topology and \textbf{distribute} the discovery functions hierarchically & Not applicable & No & Zero \\
\midrule
GTOP~\cite{choi2017design} & \textbf{Path computation element (PCE)} discovers topology as a controller & Not applicable & No & Zero\\
\bottomrule

\end{tabular}
\end{table*}

\section{Evaluation}\label{sec:eval}

We evaluate \projname{} against 10 SDN discovery protocols that use LLDP or variants for topology discovery (\S\ref{sec:eval_protocol}), and 5 open-source SDN controllers (\S\ref{sec:eval_controller}).
We demonstrate \projname{} attacks 
starting with a controller impersonation attack
on both the OpenDaylight and ONOS clusters (\S\ref{sec:eval_cluster}). We then evaluate our RL agent on two use cases: (1) traffic eavesdropping and (2) monitoring evasion (\S\ref{sec:eval_RL}). \update{Lastly, we measure \projname{}'s stealthiness against the current state-of-the-art in SDN attack detection -- PicoSDN~\cite{ujcich2021causal}.}

\subsection{Attacking discovery protocols}\label{sec:eval_protocol}

Table~\ref{table_protocols} lists various versions of SDN Discovery Protocols, including their methodology, the source MAC of the discovery packets, and whether they are based on OpenFlow.  
The first seven entries in the table aim to enhance performance by reducing unnecessary discovery packet transmissions. These protocols use customized discovery packets while relying on the same or similar discovery mechanism as OFDP, making them vulnerable to our attack. Hybrid OFDP (HOFDP) uses LLDP packets along with Broadcast Domain Discovery Protocol (BDDP) packets to discover a hybrid network (both legacy and SDN devices), following OFDP's mechanism. The BDDP packets have standard ethernet headers with \kw{ether-src}, which makes the protocol vulnerable to our poisonous flow entries.

Among the protocols listed in Table \ref{table_protocols}, only SLDP~\cite{nehra2019sldp}, TILAK~\cite{nehra2019tilak}, and sOFTP~\cite{azzouni2017softdp} address security issues related to discovery. However, they focus on the data plane, as existing topology poisoning attacks are typically initiated from the data plane. For example, SLDP randomizes the \kw{ether-src}, which inhibits a successful attack unless we use VLAN-based poisonous flow entries to regain full control. Only sOFTP provides a partial defense against \projname{} because sOFTP only partially relies on OpenFlow for link discovery.

\subsection{Attacking SDN controllers}\label{sec:eval_controller}

\begin{table}[t]
\small
\caption{\projnameTable{} attack level towards SDN Controllers}
\label{table_controllers}
\footnotesize
\setlength{\tabcolsep}{2.25pt}
\centering 
\begin{tabular}[htbp]{lllll}
\toprule
\textbf{Controller} & \textbf{Protocol} & \textbf{Security}  & \textbf{MAC LLDP/BDDP} &\textbf{Attack}\\
\midrule
Floodlight~\cite{floodlight} &  HOFDP & Hash  & src switch port &Complete \\
\midrule
OpenDaylight~\cite{opendaylight} &  OFDP & Hash  & src switch port &Complete \\
\midrule
ONOS~\cite{onos_disc} &  HOFDP & No & Fingerprints &Complete \\
\midrule
Ryu~\cite{ryu} & OFDP & No & src switch port &Complete \\
\midrule
Pox~\cite{pox} & OFDP & No & src switch port &Complete \\
\toprule
\end{tabular}
\end{table}

We evaluate 5 popular open-source SDN controllers (Table~\ref{table_controllers}) using the illustrative example introduced in \S\ref{sec:approach} assuming a topology poisoning attack initiated by a malicious application. We deploy mininet-v2.2.2, Floodlight-v1.2, OpenDaylight-v0.15.3, ONOS-2.2.0, Ryu-v4.34, and pox-v\_eel on six connected VMs with Linux systems on the CyberVAN testbed~\cite{chadha2016cybervan}. 

Among the five controllers, only OpenDaylight and Floodlight have implemented security enhancements by introducing hash checks on LLDP packets~\cite{wazirali2021sdn}. However, the hash check mechanism does not defend against \projname{} as our technique does not tamper with LLDP packets but modifies the path taken by the LLDP packets. Indeed, we achieved precise topology poisoning on all controllers. 

Interestingly, while Ryu and ONOS lack security enhancements in discovering the topology, they present additional challenges in launching a successful attack. Ryu sets its flow entries for forwarding LLDP packets back to the Ryu controller with the highest priority in the flow table.  As a result, \projname{} must lower the priority of the entries set by the Ryu controller before initiating the attack.  

ONOS employs per-logical-plane fingerprints encoded on the \kw{ether-src} field, resulting in a unified value of \kw{ether-src} per cluster. As a result, VLAN-based poisonous flow entry matching [\kw{ether-src}, \kw{in-port}] is mandatory to identify the source of discovery packets for all three deceptive links. Details are described in Section \ref{sec:vlanentries}. \update{To conclude, \projname{} can attack all 5 open-source controllers. }

\inlinedsection{Discussion}
Discovery packet flow entries are static and specifically serve the control plane. As a result, they should not be changed or overridden easily by arbitrary controllers. Unfortunately, there is currently no access control mechanism for flow entry enforcement implemented in any of the SDN controllers examined.

\subsection{Attacking SDN Controller Clusters}\label{sec:eval_cluster}

\begin{figure}[t]
\centering
\subfigure[Original topology on ONOS]{\label{fig:onos_realTopo} 
\includegraphics[width=0.47\textwidth]
{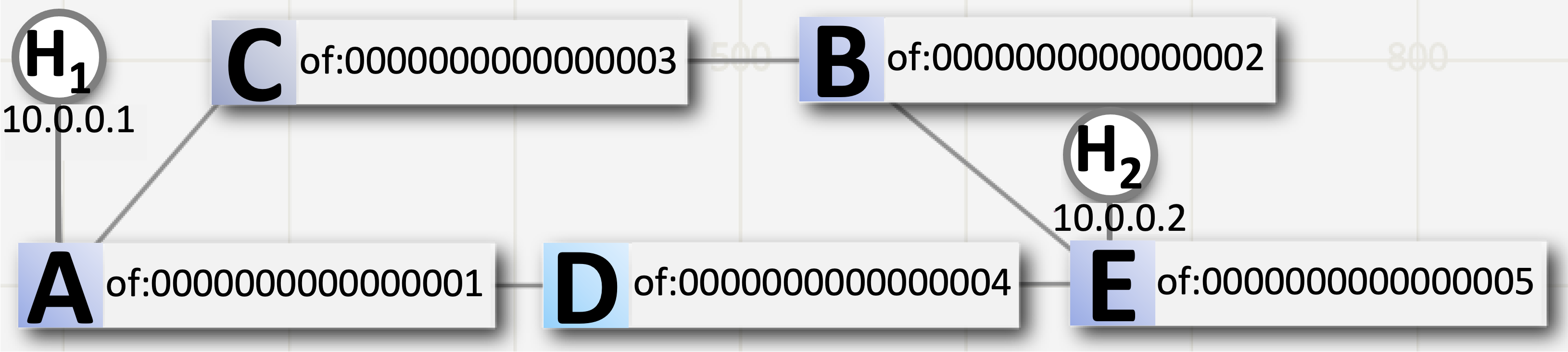}}
\hfill
\subfigure[Topology after attack on ONOS]{\label{fig:onos_fabTopo} \includegraphics[width=0.47\textwidth]{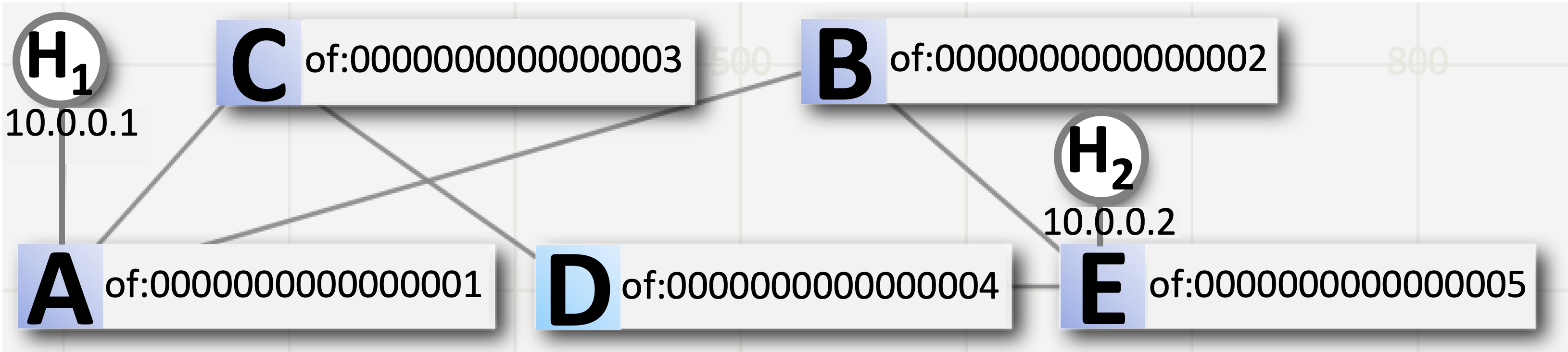}}%
\caption{\projnameTable{} attack on ONOS cluster}
\label{fig:eval_onos_cluster}
\vspace{-10pt}
\end{figure}
Among all open-source controllers, only OpenDaylight and ONOS support controller clustering~\cite{opendaylight}~\cite{onos_disc}. To assess the attack in both OpenDaylight and ONOS cluster scenarios, we first initiate a controller impersonation attack (Appendix \ref{sec:eval_impersonation}) occurring when a controller goes offline, and a malicious machine located in the same subnet exploits the opportunity to join the cluster by impersonating the offline controller. The experiment on the ONOS cluster is described below; the attack on OpenDaylight is similar.

We construct a three-node ONOS cluster. The cluster is built with ONOS docker \texttt{onosproject/onos:2.2.2} and atomic docker \texttt{atomix/atomix:3.1.5} on the host machine with Ubuntu 20.04.6 LTS system. Its network is simulated with \texttt{mininet-v2.3.0} on the host machine. The third controller ONOS-3 is in passive replication mode without a direct OpenFlow connection to the mininet.

We recreate our example scenario (\S\ref{sec:approach})
on the ONOS cluster and the results of the attack are shown in Figure \ref{fig:eval_onos_cluster}. 
We use VLAN-based poisonous flow entries to poison the topology view of the controller cluster.

In this experiment, we define a topology on mininet as shown in Figure \ref{fig:onos_realTopo} and connect it with ONOS-1 and ONOS-2; ONOS-3 is not connected, \update{serving as a passive replication of the control plane. The malicious ONOS-mal takes ONOS-3's identity in the cluster and sets up computed poisonous flow entries regardless of not connecting with mininet, either. After that, the topology view changes on all the controllers and settles as shown in Figure \ref{fig:onos_fabTopo}.} Nodes B and D switched their locations.

ONOS has reactive forwarding enabled so it fixes the gap automatically when we start the ping from $H_1$ to $H_2$ after the topology poisoning. Because the version of ODL we use does not have an available forwarding application, we wrote a proactive flow rule installation application on it. To show the gap, we only sent gap-patching flow entries after the controller changed the routing based on the deceptive topology. 

\inlinedsection{Discussion}In both OpenDaylight and ONOS clusters, any controller, regardless of role, can set up flow entries even without a direct OpenFlow connection due to the data store consistency mechanism. When a passive replication controller sets up flow entries, it actually updates its flow table data store. This update is captured and accepted without evaluation by other controllers including the leaders. As a result, the leader controllers set up the flow entries to the network on behalf of it. 

\subsection{Poisonous Topology Computation by RL}\label{sec:eval_RL}

Enterprise networks and backbone networks are both common SDN scenarios~\cite{li2019application,nencioni2017impact}. To validate the practicality of our attack, we evaluate our RL algorithm on an enterprise \textit{fat tree} topology~\cite{Leiserson:1985} and 
a backbone network with centralized nodes named \textit{Chinanet}~\cite{topozoo}.  
Table~\ref{table:RL_topo_flow_info} summarizes the topologies and flows used in our evaluation. 

We implemented the RL algorithm with Python3.9 using the Stable-Baselines3~\cite{stablebaselines}. The 2-switch and node reallocation actions are used
because these are the unit actions to change a graph while maintaining the same degree sequence, and they are efficient actions for training on the tree topology and cyclic topology, respectively.
Because evading monitoring is harder to achieve, we set 10 actions per episode for Chinanet case and 5 for FatTree. With the equation of $n_a = n(1-s)$, where $n$ is the number of links, $s$ is the similarity, and $n_a$ is the number of actions per episode (\S\ref{sec:design_RL}), the similarity threshold for FatTree and Chinanet are roughly set to 0.9 and 0.8, respectively.
For each step, a reward of 1 is granted to the RL agent if: (1) the number of the shortest path routes covered by the eavesdropping/monitoring node is greater/smaller than the flow coverage goal and (2) the similarity is above the threshold value. Otherwise, a -1 is granted. 
Because a failed action causes a -1 reward, the acceptable Episode Reward Mean of FatTree and Chinanet topologies are the ones larger than -5 and -10, respectively, meaning at least one success is made.

\begin{table}[t]
\small
\caption{Networking Scenario and Randomized Flows}
\footnotesize
\label{table:RL_topo_flow_info}
\centering 
\setlength{\tabcolsep}{2pt}
\begin{tabular}{llcccc}
\toprule
\textbf{Attack Goal} & \textbf{Topology} & \textbf{\# Nodes} & \textbf{\# Edges} & \textbf{\# Flows} & \textbf{Max Degree}\\
\midrule
Eavesdrop Node 6 & FatTree & 36 & 48 & 120 & 4\\
\midrule
Evade Node 8 & Chinanet & 42 &66 & 89  & 20\\
\bottomrule
\end{tabular}
\end{table}

\begin{figure}[t!]
    \centering
     \subfigure[Real Fat Tree with original routes]
        {\includegraphics[width=0.15\textwidth]{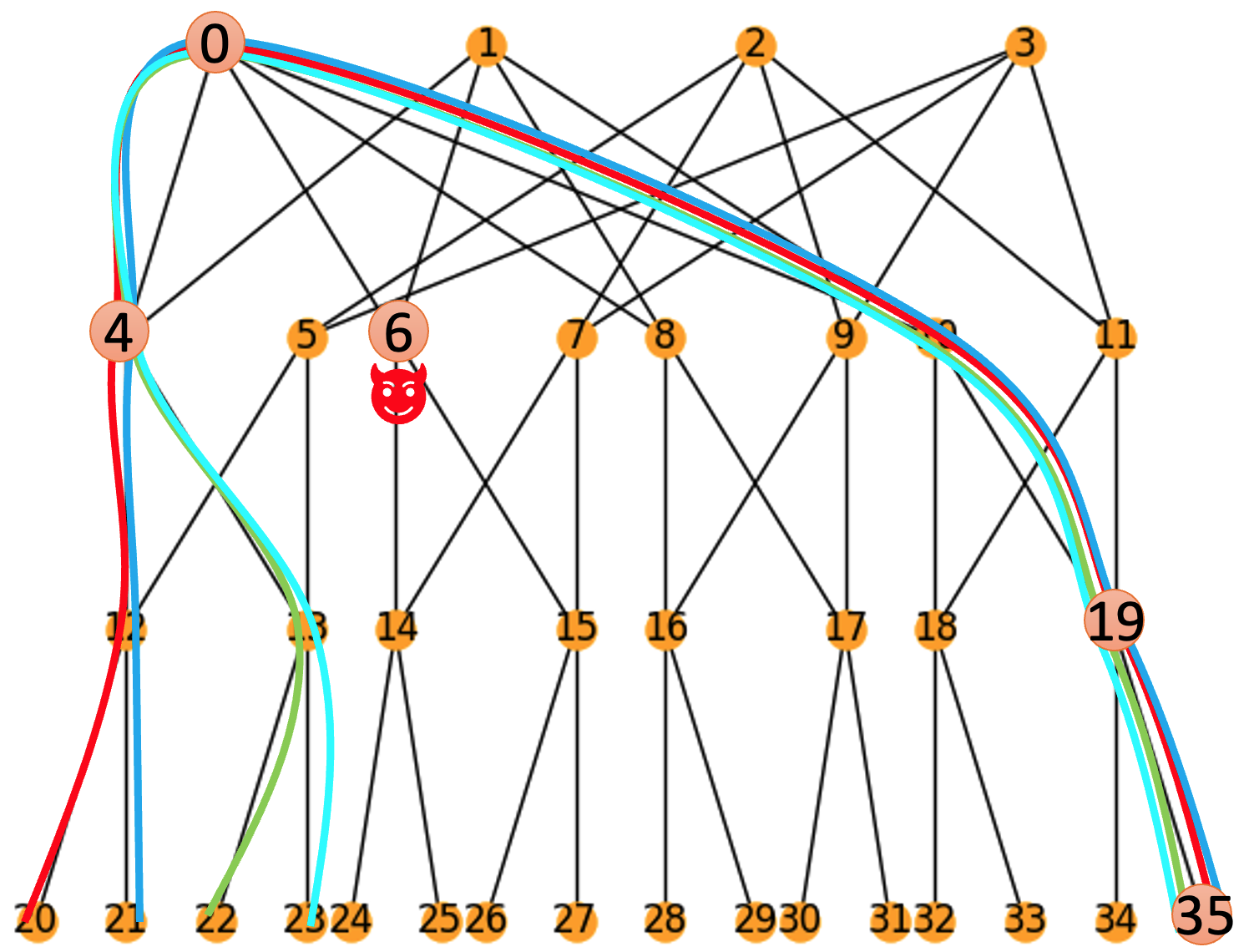}\label{fig:fat_tree_real_topo}}
     \hfill
     \subfigure[Deceptive Fat Tree with updated routes]
        {\includegraphics[width=0.15\textwidth]{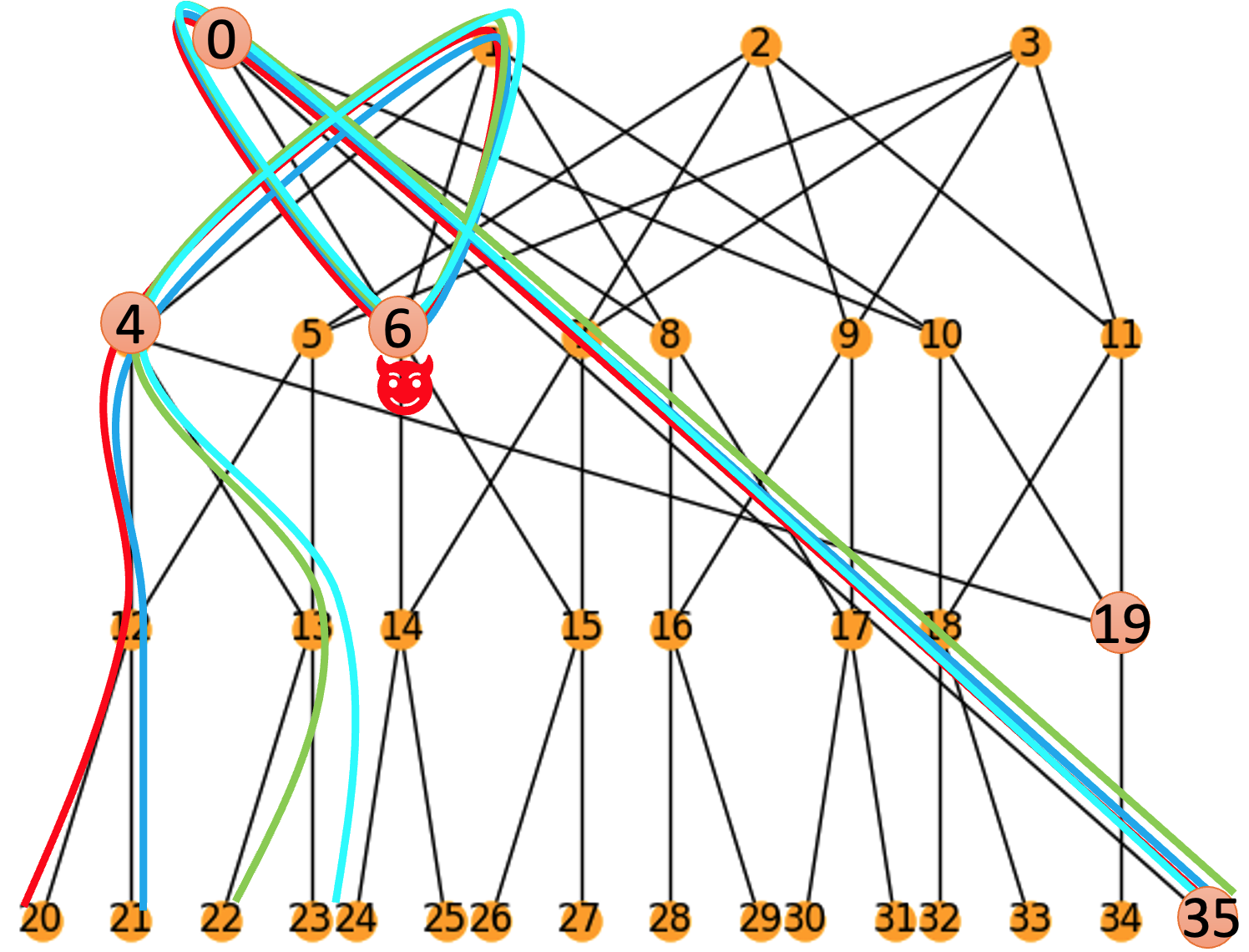}\label{fig:fattree_dec_topo}}
    \hfill
    \subfigure[Real Fat Tree with updated routes]
        {\includegraphics[width=0.15\textwidth]{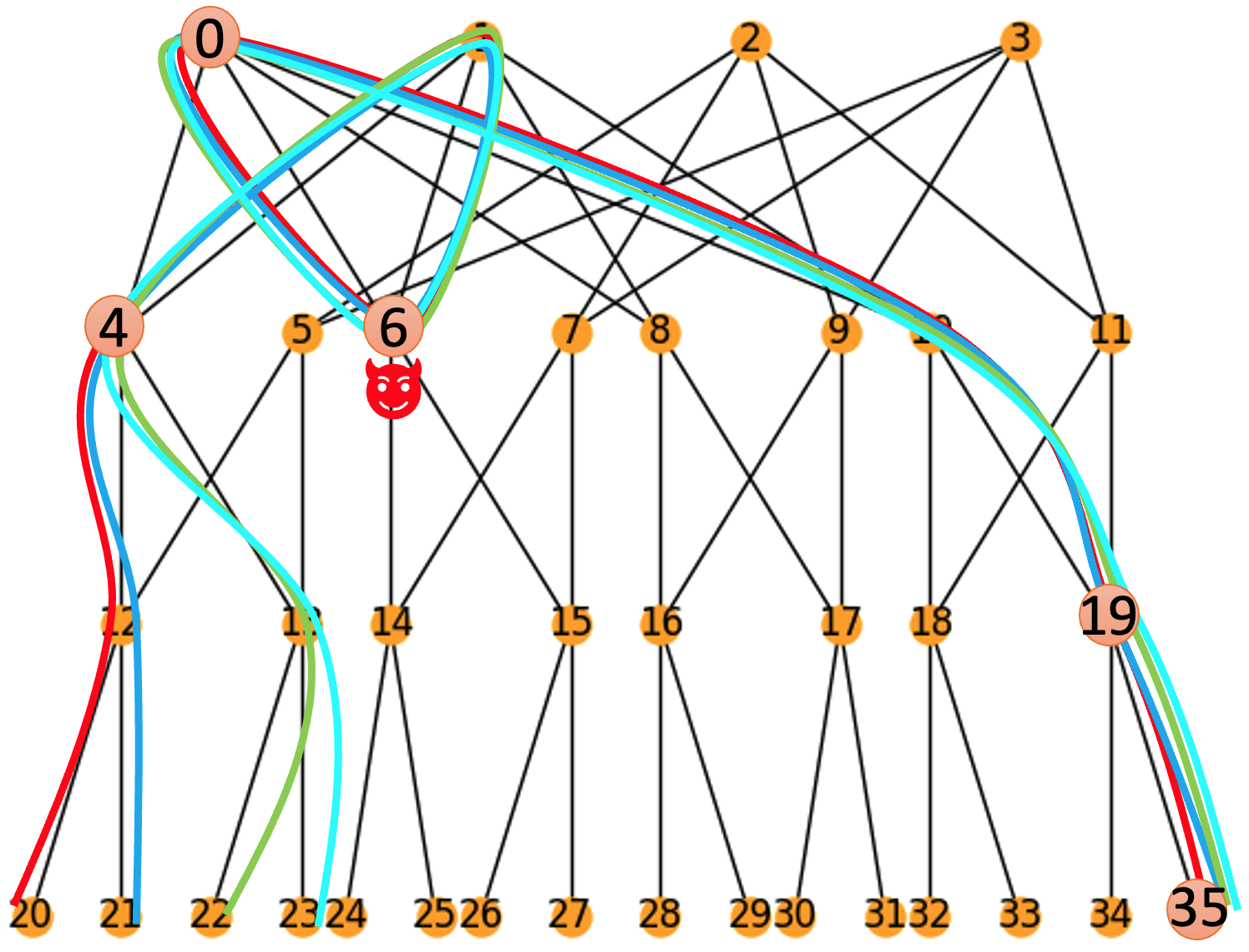}\label{fig:fattree_real_topo_upd_r}}
    \caption{Eavesdrop Node 6 (Fat Tree) covering 4 extra flows}
    \label{fig:eaves_FatTree_topo}
     \vspace{-10pt}
\end{figure}

\inlinedsection{Eavesdropping on Enterprise Network}\label{sec:eval_eaves_RL}
We depict a general fat tree topology that is used as our enterprise network. The goal of this example (Figure \ref{fig:eaves_FatTree_topo}) is to determine a poisonous topology that will route four additional flows through Node 6 using shortest path routing. Node 6 has 52 out of 120 randomly generated flows covered originally. We use the 2-switch action to train the model since this is a tree-like topology.

\begin{figure}[t!]
    \centering
     \subfigure[RL achieved metrics]
        {\includegraphics[width=0.245\textwidth]{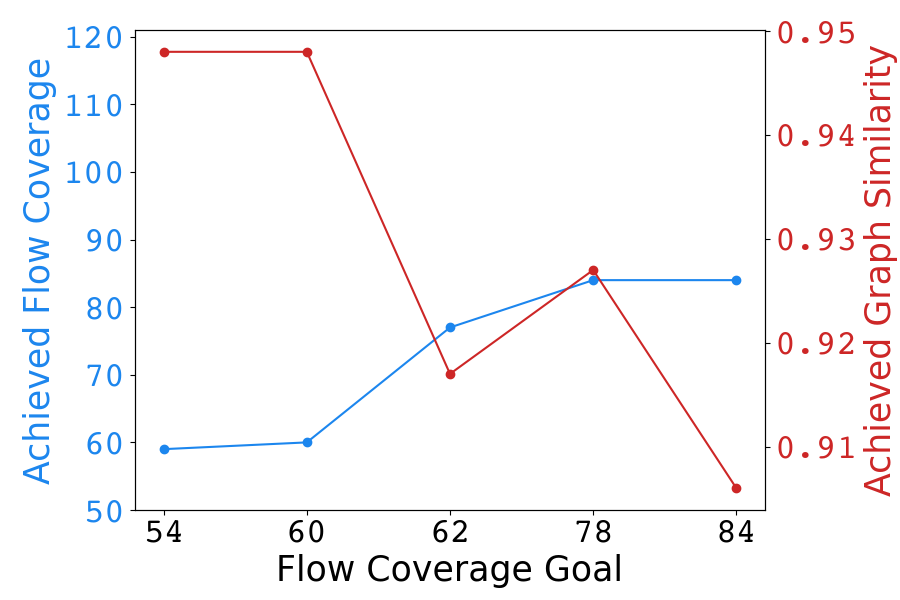}\label{fig:Eveas_res}}
     \hfill
     \subfigure[RL training log]
        {\includegraphics[width=0.225\textwidth]{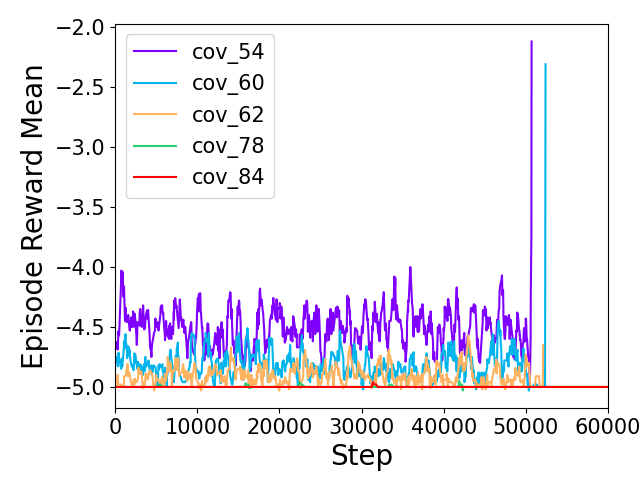}\label{fig:Eveas_training_log}}
    \caption{RL for different eavesdropping goals (Fat Tree)}
    \label{fig:RL_FatTree}
    \vspace{-10pt}
\end{figure}

Figure~\ref{fig:fat_tree_real_topo} shows the real topology with the four flows on their original routes. Figure \ref{fig:fattree_dec_topo} shows the deceptive topology computed by RL. The routes of the four flows are changed based on the deceptive topology. They all traverse Node 6 according to the shortest path routing. In Figure \ref{fig:fattree_real_topo_upd_r}, the induced routes of those four flows still traverse Node 6 on the real topology, thus achieving the goal. The induced shortest path routes of those four flows have two gaps, one at Node 10 and one at Node 19, that will be patched by the gap patching module discussed in \S\ref{sec:gap_patching}. 

Figure \ref{fig:RL_FatTree} shows the details of the RL algorithm results for 5 different flow coverage goals and their corresponding episode reward mean (ERM) during the training. In Figure \ref{fig:Eveas_res}, the graph similarity decreases with an increase in the flow coverage goal which means more changes are required to eavesdrop on more flows. We achieved up to 32 additional flows being routed through Node 6 as a result of the deceptive topology while still maintaining the same degree sequence as the original topology and achieving a similarity larger than 0.9 when compared with the real topology. 

Figure \ref{fig:Eveas_training_log} shows that the ERM of the three lowest coverage goals stops at around 50,000 steps (indicating a convergence) because the callback function to stop training when the reward meets threshold -4 is set. However, the ERM for covering 84 flows at Node 6 fails to converge (maintained -5 except for 1 timepoint) and never stops before the end of the total steps, meaning this is a hard goal to achieve.  However, the RL still found a solution because of the 1 timepoint (around Step 30000) with ERM greater than -5.

\begin{figure}[t]
    \centering
     \subfigure[Real Chinanet with original routes]
        {\includegraphics[width=0.15\textwidth]{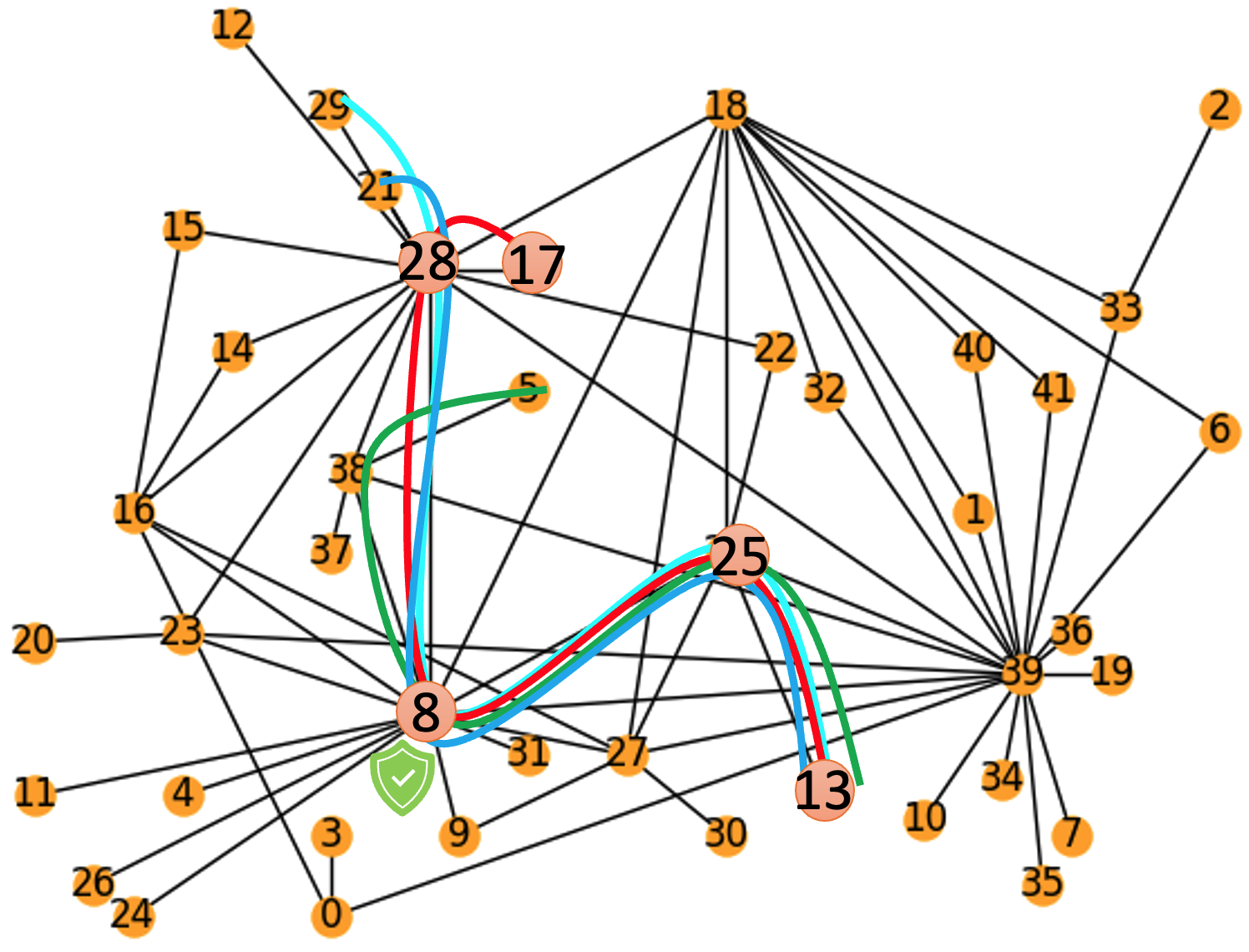}\label{fig:Chinanet_real_topo}}
     \hfill
     \subfigure[Deceptive Chinanet with updated routes]
        {\includegraphics[width=0.15\textwidth]{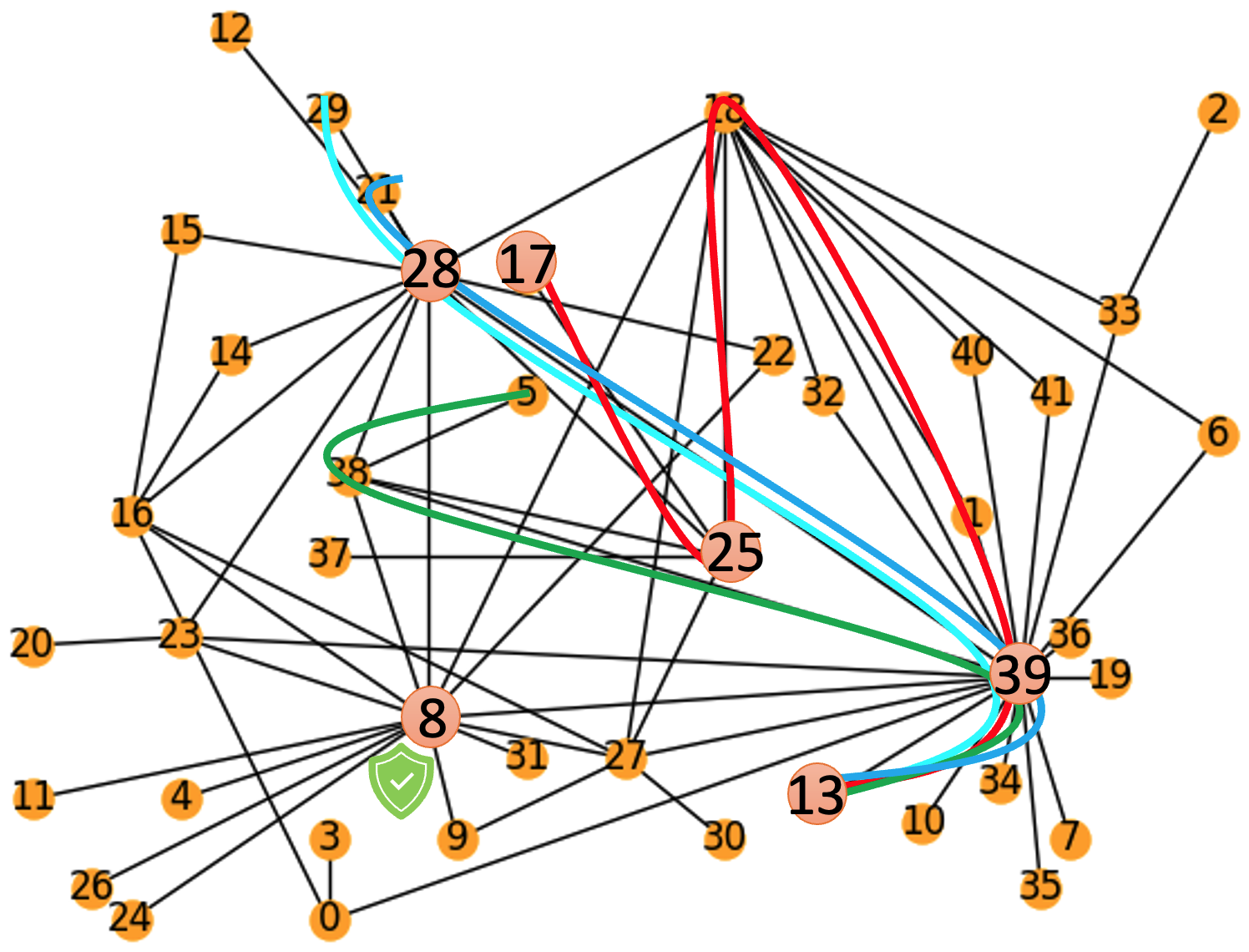}\label{fig:Chinanet_dec_topo}}
     \hfill
     \subfigure[Real Chinanet with updated routes]
        {\includegraphics[width=0.15\textwidth]{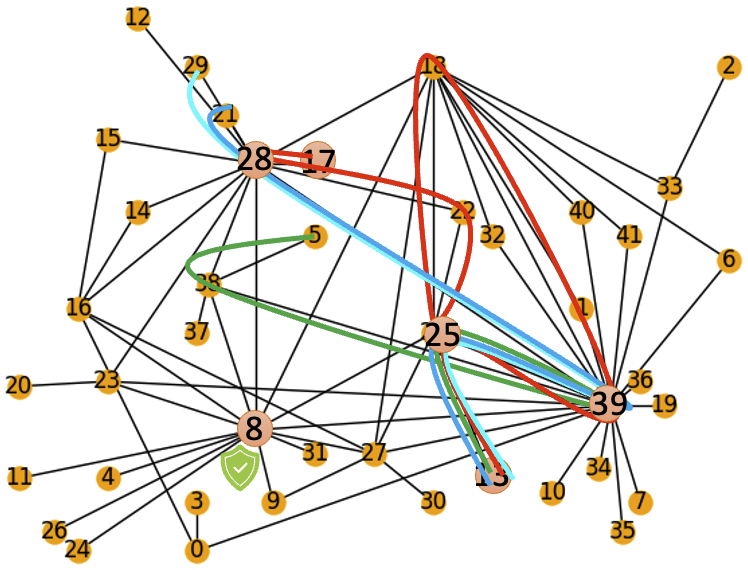}\label{fig:Chinanet_real_topo_upd_r}}
    \caption{Goal: Remove 4 flows from Node 8 (Chinanet)}
    \label{fig:pois_Chinanet_topo}
    \vspace{-10pt}
\end{figure}

\inlinedsection{Evading Monitoring on Backbone Network}\label{sec:eval_mon_RL}
The Chinanet topology is shown in Figure \ref{fig:pois_Chinanet_topo}.  The attack goal here is to drive four flows away from Node 8 which may be a monitoring point that we wish to avoid.  Node 8 originally covers 51 out of 89 randomly generated flows. We use node-reallocation action to train the model since it is a mesh-like topology.

\begin{figure}[t]
    \centering
     \subfigure[RL achieved metrics]
        {\includegraphics[width=0.245\textwidth]{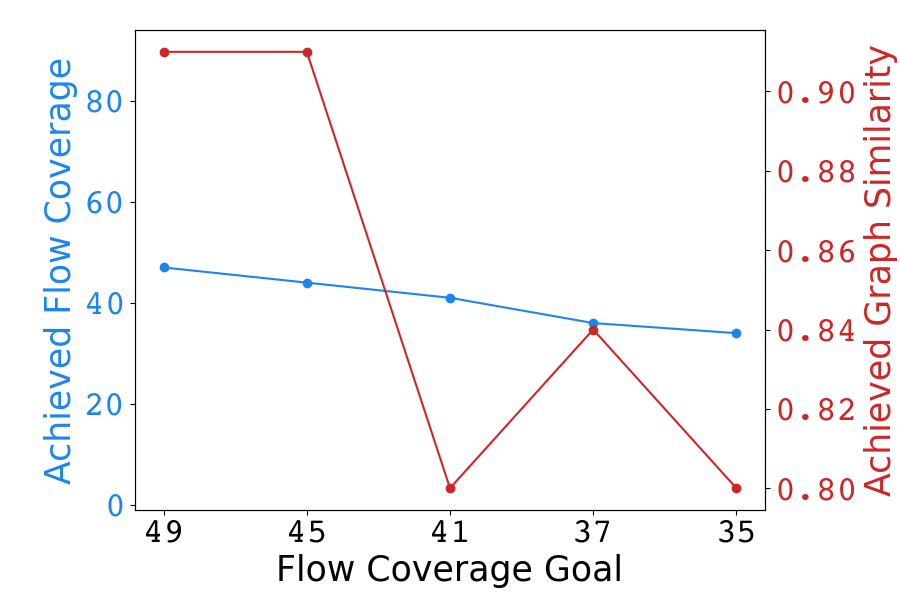}\label{fig:EvaMon_res}}
     \hfill
     \subfigure[RL training log]
        {\includegraphics[width=0.225\textwidth]{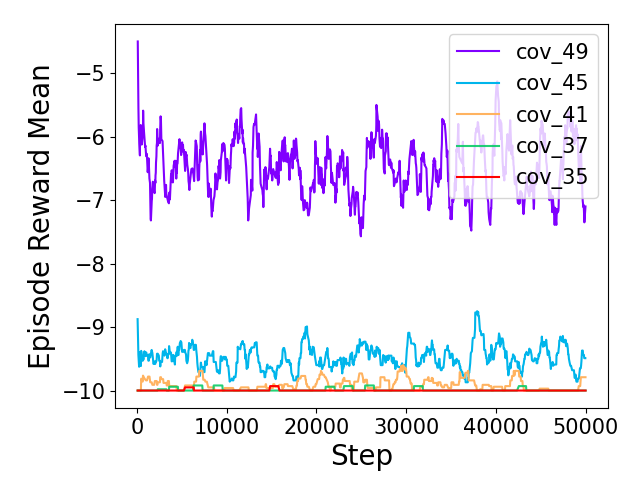}\label{fig:EvaMon_log}}
    \caption{RL for different evasion goals (Chinanet)}
    \label{fig:RL_Chinanet}
    \vspace{-10pt}
\end{figure}

Figure \ref{fig:Chinanet_real_topo} shows the real topology with the four flows on their original routes, all traversing Node 8. Figure \ref{fig:Chinanet_dec_topo} shows the deceptive topology learned by our RL model.  The routes of the four flows are changed based on the deceptive topology and no longer go through Node 8. Instead, as shown in Figure~\ref{fig:Chinanet_real_topo_upd_r}, they all go through Node 39.
Figure \ref{fig:RL_Chinanet} shows the details of the RL algorithm results and the ERM for five different goals for evading Node 8. When the flow coverage is smaller than our flow coverage upper bound, we grant a reward of 1. Otherwise, the reward is -1. 
 
\update{Figure \ref{fig:RL_Chinanet}} shows the RL results and corresponding ERM during the training of 5 different flow coverage goals.  We successfully divert up to 17 flows away from being monitored at Node 8 by computing
a deceptive topology with the same degree sequence as the original topology and a similarity score larger than 0.8 when compared with the real topology. Figure \ref{fig:EvaMon_res} shows that the graph similarity generally decreases with the decreased expected flow coverage upper bound which means more changes are required to drive more flows away from being monitored. 
Figure \ref{fig:EvaMon_log} shows that the ERM values never converge. However, the ERM is higher when the training goal is easier which means fewer actions are needed to achieve the goal in each episode. Because we do not target training the model to converge but just a one-time solution, as long as the ERM is greater than -10, the RL has found a solution during the training. 

\inlinedsection{Discussion}
The poisonous topology computation by RL does not have scalability issues because the deceptive topology is computed offline and we do not poison the topology frequently to keep the attack stealthy. We do not need a stable agent returning a solution with real-time flows as input. Instead, we observe and select representative flows as input for the RL training. During the training, any reward greater than the minimum ERM means a deceptive topology meeting our goal has been captured. Moreover, the topology only has hundreds of nodes, and changes should correspond to a small segment (e.g., around the eavesdropping node) of the large production networks.

\update{To validate the RL acceleration of action priors technique in Section \ref{sec:action_prior}, we also used a 2-switch action to train the model to evade monitoring on the Chinanet topology to compare it with the node reallocation. In this case, ERM remains -10 for all 50,000 steps even with the simplest goal of driving one flow away from Node 8. Intuitively, that is because a single 2-switch action cannot achieve the goal of evading a node on the mesh-like Chinanet topology and the exploration space is huge ${66\choose2}^{10}$.
On the other hand, the node reallocation action returns a successful result quickly because very few node reallocations are necessary on the cyclic Chinanet topology to change the length of some paths leading to driving traffic away from a node. 
}

\subsection{Evading Existing Detection}\label{sec:eval_defense}
\begin{figure*}
    \centering
    \includegraphics[width=0.9\textwidth]{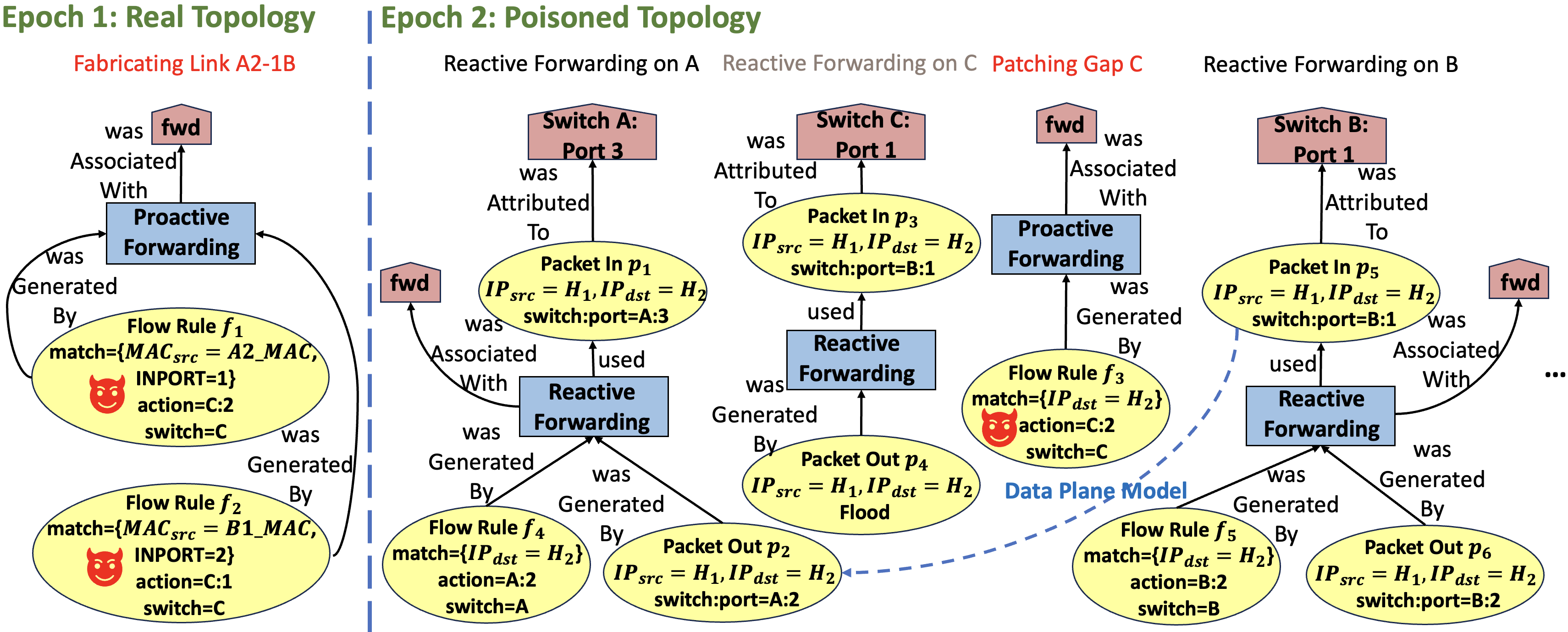}
    \caption{Relevant provenance for the \projnameTable{} attack. PicoSDN failed to detect poisonous flow entries $f_1,f_2$ and gap-patching flow entry $f_3$.}
    \label{fig:picoSDN}
\end{figure*}

(D.7) Monitor-based detection is the only detection relevant to \projname{}. 
Among all the systems listed in (D.7), only the code of PicoSDN has been made available. The other types of defenses described in the paper focus on characteristics of the network that are not altered by \projname{}, so they cannot detect it by design. These characteristics include degree sequence, LLDP packet integrity, information integrity, and traffic routing integrity. Note that because there are defenses (D.3-5) that measure degree sequence, one of the core components of \projname{} is to ensure the degree sequence remains unchanged. (D.8) Voting-based detection can address spoof-based attacks and suspicious activity involving multiple entities with respect to decision-making (e.g. reactive forwarding decisions among controllers), but the \projname{} does not spoof; it unilaterally inserts malicious flow entries with no other controllers involved to evaluate its behavior.

\update{We have comprehensively analyzed the existing related defenses (Table \ref{table:related_threat}) and their ineffectiveness against \projname{} (Table \ref{table:features}). In this section, we provide an in-depth evaluation of PicoSDN~\cite{ujcich2021causal}. }
PicoSDN~\cite{ujcich2021causal} is a provenance-informed monitoring-based detection system that analyzes the logs of network events. 
PicoSDN utilizes a data plane model that provides topology information enabling packet tracing on the data plane, combined with the trace of the processing of the packet on the control plane. This facilitates the construction of a comprehensive packet trace throughout the network (on both data and control planes) to reveal a fine-grained causal analysis.

The GitHub repository of PicoSDN~\cite{picoSDN_github} does not provide any algorithm implementation to build the provenance graph, but only the classes and methods to operate on a provenance graph (confirmed with the author). Thus we manually construct a snippet of the provenance graphs of fabricating link $A2\rightarrow1B$ in Figure \ref{fig:motivation_example} and the consequent gap patching for traffic from $H_1$ to $H_2$  according to the algorithm presented in the PicoSDN~\cite{ujcich2021causal} as shown in Figure \ref{fig:picoSDN}.

PicoSDN fails to detect the topology poisoning attack of fabricating link $A2\rightarrow1B$ caused by poisonous flow entries $f_1, f_2$ of \projname{} because it overlooks that flow entries can attack the topology discovery result. Whenever the topology is changed, either legitimately or through a \projname{} attack, PicoSDN starts a new epoch to construct a provenance graph independently~[Algorithm 1(line 1-2)~\cite{ujcich2021causal}]. In Epoch 1, \projname{} attacks the topology view based on real topology. As a result of the poisonous flow entries, PicoSDN starts Epoch 2 due to the changed topology.

PicoSDN fails to detect the gap patching flow entries $f_3$ because it fails to recognize that Switch $C$ is between Switch $A$ and Switch $B$ due to the undetected poisoned topology. When traffic starts from $H_1$ to $H_2$, PicoSDN constructs a data plane model based on the deceptive link of $A1\rightarrow1B$ so it links Packet-Out $p_2$ at $A$ with Packet-In $p_5$ at $B$~[Algorithm 1(line 7-10)~\cite{ujcich2021causal}] which misses Packet-In $p_3$ and Packet-Out $p_4$ at $C$ in the gap~\footnote{The controller \emph{flood}s the $p_4$ due to not knowing a path to forward it on the altered topology, so $B$ can still receive $p_5$.}. Consequently, the gap patching flow entry $f_3$, set to prevent unnecessary flooding of reactive forwarding, is not linked to the traffic from $H_1$ to $H_2$, evading the anomaly detection successfully. We omit the Reactive Forwarding on $E$ which is in the same pattern as Reactive Forwarding on $B$.  To conclude, the causal analysis by PicoSDN based on its fine-grained provenance graph fails to detect \projname{}.

\inlinedsection{Discussion}
PicoSDN and existing related defenses overlook the vulnerability of the SDN topology discovery process that certain flow entries can impact the link discovery outcome. The detection of this attack is impossible without monitoring flow rule conflicts against table-miss/LLDP flow entries, which has not been studied. Failing to detect the control plane topology poisoning is troublesome. A direct consequence is that the gap patching flow entries also evade PicoSDN's detection because causal analysis based on a false topology fails to link the gap patching flow entries with the flows that these gap patching flow entries are patching. 

\section{Discussion}\label{sec:discussion}

\inlinedsectionit[0pt]{Defending against \projname{}} 
\update{
Detecting a \projname{} attack involves recognizing that flow entries intended for traffic forwarding can fabricate links. Therefore, monitoring flow rule conflicts on table-miss/LLDP flow entries becomes crucial. } More importantly, the discovery protocol should be included in the OpenFlow protocol so that it can be regulated and implemented to a set of standards.  

Monitoring-based detection is an option to detect \projname{}'s attack if the attack signature is known. However, augmenting a monitor on the control plane is cumbersome and the monitoring-based detection can only detect the attack, not defend it. It is more effective to make slight modifications to the discovery packets or discovery process to secure the link discovery result. 

According to the OpenFlow protocol, if a packet with an invalid TTL is received at a switch, this switch must drop the packet or send it to the controller. A naive defense against \projname{} is to make \kw{TTL=1} for the discovery packet to force the switch to send it back to the controller instead of allowing it to be forwarded. However, this method is not robust for two reasons: (1) the \kw{TTL} implementation on OpenFlow switches may vary; we found that the OpenVSwitch in Mininet will drop the packet with an invalid TTL only when it has a corresponding flow entry that has an action of decreasing the \kw{TTL} (\kw{dec\_nw\_ttl}), otherwise, this packet will still be forwarded in Mininet environment. The process of confirming \kw{TTL==0} on \kw{dec\_ttl} action is specified in the Open vSwitch Manual~\cite{ovs}. As a result, this process may also apply to some physical OpenFlow switches because white box OpenFlow switches are just x86 machines running OpenVSwitch with some variants. For example, the OpenFlow switches with the PICOS system by Pica8 have a default \kw{dec\_ttl} action but it can be overwritten by ECMP select group flow or \kw{set-l3-egress-keep-fields}~\cite{pica8_manual2}. They also depend on flow entries with the action of \kw{to\_controller} (\kw{reason} = \kw{invalid\_ttl}) to send the invalid-TTL packet to the controller~\cite{pica8_manual1}. All of the above can be manipulated by \projname{}. (2) With the ability to set flow entries, the \projname{} can set flow entries by setting the \kw{TTL} (\kw{set\_nw\_ttl}) to make the discovery packet’s TTL valid again. 

Although the LLDP has been widely used by SDN to discover links, the LLDP packet is not designed for SDN but for the traditional layer 2 networks. As a result, many fields of the LLDP payload are redundant and expose vulnerabilities.
Our patching plan is to let OpenDaylight randomize the \kw{ether-src}, \kw{ether-dst}, and \kw{ether-type} of the discovery packets each time they are sent. Thus, there will be no way for a flow entry to match these discovery packets, guaranteeing that they are a miss and will be sent back to the controller due to the default table-miss flow entry. This method brings complexity to the controllers because the controller needs to maintain and update a table of matching the randomized \kw{ether-src} with the real \kw{ether-src} to know the source of the received discovery packet. The discovery packet’s destination can be known from the \kw{packet-in} message without any changes. Because the table needs to be updated with the frequency of the topology discovery process, maintaining its data consistency is nontrivial.

\inlinedsectionit{Limitations} 
In a single-controller setting, \projname{} may need to frequently query the SDN flow table to patch the gaps for long-lasting flows, potentially raising suspicion. However, in a controller cluster setup, all controllers get flow table update notifications, eliminating the need for querying the flow table. Moreover, hit rate statistics on original flow entries
\update{will not be kept consistent when it is mandatory to use an LLDP signature field (e.g. \kw{ether-src}) combined with \kw{in-port} to distinguish LLDP packets from different sources to fabricate links precisely.}

\inlinedsectionit{Ethical disclosure and open-sourcing} We have gone through an ethical disclosure with the OpenDaylight maintainers, who acknowledged the vulnerability and CVE-2024-37018 has been assigned. We will patch this vulnerability and contribute to the OpenDaylight source code. We have published the source code with documentation to reproduce our attack work~
\cite{Chen2024-to}.

\section{Conclusion}\label{sec:conclusion}

We describe a new SDN link fabrication attack that is global, stealthy, and persistent.  Launched from the control plane, a single compromised controller \update{ or a malicious application} can manipulate all other controllers in a cluster into learning a poisoned topology by influencing the paths of link layer discovery packets.  Link fabrication attacks can route traffic in nefarious ways, to eavesdrop on a set of devices, or avoid a network monitoring device.  

To scale the attack to large networks, we present a framework based on reinforcement learning called \projname{}, which, given the network topology, and an attacker goal, can automatically generate the set of poisonous OpenFlow flow table entries required to launch the attack.  Results show that our approach can successfully attack 9 discovery protocols, and 5 controllers we tested, including controller clusters. Our fabricated links also go undetected by current state-of-the-art defenses.  

\section{Acknowledgements}
We thank the anonymous reviewers and shepherd for their valuable comments and constructive feedback that enhanced this paper. 
This research was sponsored by the U.S. Army Combat Capabilities Development Command Army Research Laboratory and was accomplished under Cooperative Agreement Number W911NF-13-2-0045 (ARL Cyber Security CRA). The views and conclusions contained in this document are those of the authors and should not be interpreted as representing the official policies, either expressed or implied, of the Combat Capabilities Development Command Army Research Laboratory of the U.S. government. The U.S. government is authorized to reproduce and distribute reprints for government purposes notwithstanding any copyright notation here on.

\bibliographystyle{plain}
\bibliography{ref}

\begin{thebibliography}{10}

\bibitem{ryu}
Build sdn agilely.
\newblock \url{https://ryu-sdn.org/}, 2011.
\newblock Accessed on 2023-06-19.

\bibitem{abdou2018comparative}
AbdelRahman Abdou, Paul~C Van~Oorschot, and Tao Wan.
\newblock Comparative analysis of control plane security of sdn and conventional networks.
\newblock {\em IEEE Communications Surveys \& Tutorials}, 20(4):3542--3559, 2018.

\bibitem{adjou2022topotrust}
Mohamed~Lamine Adjou, Chafika Benza{\"\i}d, and Tarik Taleb.
\newblock Topotrust: A blockchain-based trustless and secure topology discovery in sdns.
\newblock In {\em 2022 International Wireless Communications and Mobile Computing (IWCMC)}, pages 1107--1112. IEEE, 2022.

\bibitem{ahmad2015security}
Ijaz Ahmad, Suneth Namal, Mika Ylianttila, and Andrei Gurtov.
\newblock Security in software defined networks: A survey.
\newblock {\em IEEE Communications Surveys \& Tutorials}, 17(4):2317--2346, 2015.

\bibitem{al2022link}
Ismail Al~Salti and Ning Zhang.
\newblock Link-guard: an effective and scalable security framework for link discovery in sdn networks.
\newblock {\em IEEE Access}, 10:130233--130252, 2022.

\bibitem{al2010flowchecker}
Ehab Al-Shaer and Saeed Al-Haj.
\newblock Flowchecker: Configuration analysis and verification of federated openflow infrastructures.
\newblock In {\em Proceedings of the 3rd ACM workshop on Assurable and usable security configuration}, pages 37--44, 2010.

\bibitem{alharbi2015security}
Talal Alharbi, Marius Portmann, and Farzaneh Pakzad.
\newblock The (in) security of topology discovery in software defined networks.
\newblock In {\em 2015 IEEE 40th Conference on Local Computer Networks (LCN)}, pages 502--505. IEEE, 2015.

\bibitem{alimohammadifar2018stealthy}
Amir Alimohammadifar, Suryadipta Majumdar, Taous Madi, Yosr Jarraya, Makan Pourzandi, Lingyu Wang, and Mourad Debbabi.
\newblock Stealthy probing-based verification (spv): An active approach to defending software defined networks against topology poisoning attacks.
\newblock In {\em European Symposium on Research in Computer Security}, pages 463--484. Springer, 2018.

\bibitem{alvarez2020hddp}
Joaquin Alvarez-Horcajo, Elisa Rojas, Isaias Martinez-Yelmo, Marco Savi, and Diego Lopez-Pajares.
\newblock Hddp: Hybrid domain discovery protocol for heterogeneous devices in sdn.
\newblock {\em IEEE Communications Letters}, 24(8):1655--1659, 2020.

\bibitem{unauthenticated_access_cisco}
Liviu ARSENE.
\newblock Apic vulnerability in cisco`s sdn controller allows unauthenticated remote root access.
\newblock \url{https://www.bitdefender.com/blog/hotforsecurity/apic-vulnerability-in-ciscos-sdn-controller-allows-unauthenticated-remote-root-access/}, 2015.

\bibitem{onos_disc}
Shaoyong~Wu Ayaka~Koshibe.
\newblock Onos network discovery.
\newblock \url{https://wiki.onosproject.org/display/ONOS/Network+Discovery}, 2016.
\newblock Accessed on 2023-05-25.

\bibitem{azzouni2017softdp}
Abdelhadi Azzouni, Raouf Boutaba, Nguyen Thi~Mai Trang, and Guy Pujolle.
\newblock softdp: Secure and efficient topology discovery protocol for sdn.
\newblock {\em arXiv preprint arXiv:1705.04527}, 2017.

\bibitem{baidya2020link}
Sonali~Sen Baidya and Rattikorn Hewett.
\newblock Link discovery attacks in software-defined networks: Topology poisoning and impact analysis.
\newblock {\em J. Commun.}, 15(8):596--606, 2020.

\bibitem{barrus20122}
Michael~D Barrus.
\newblock On 2-switches and isomorphism classes.
\newblock {\em Discrete Mathematics}, 312(15):2217--2222, 2012.

\bibitem{stablebaselines}
Stable Baselines3.
\newblock Stable-baselines3 docs - reliable reinforcement learning implementations.
\newblock \url{https://stable-baselines3.readthedocs.io/en/master/}, 2022.

\bibitem{picoSDN_github}
Samuel~Jero Benjamin E.~Ujcich.
\newblock Provenance for software-defined networking (sdn).
\newblock \url{https://github.com/bujcich/PicoSDN/tree/main}, 2021.

\bibitem{cao2020match}
Jiahao Cao, Renjie Xie, Kun Sun, Qi~Li, Guofei Gu, and Mingwei Xu.
\newblock When match fields do not need to match: Buffered packets hijacking in sdn.
\newblock In {\em Proc. of the Network and Distributed System Security Symposium (NDSS'20)}, 2020.

\bibitem{chadha2016cybervan}
Ritu Chadha, Thomas Bowen, Cho-Yu~J Chiang, Yitzchak~M Gottlieb, Alex Poylisher, Angello Sapello, Constantin Serban, Shridatt Sugrim, Gary Walther, Lisa~M Marvel, et~al.
\newblock Cybervan: A cyber security virtual assured network testbed.
\newblock In {\em MILCOM 2016-2016 IEEE Military Communications Conference}, pages 1125--1130. IEEE, 2016.

\bibitem{chang2018fast}
Sang-Yoon Chang, Younghee Park, and Bhavana Babu~Ashok Babu.
\newblock Fast ip hopping randomization to secure hop-by-hop access in sdn.
\newblock {\em IEEE Transactions on Network and Service Management}, 16(1):308--320, 2018.

\bibitem{Chen2024-to}
Mingming Chen.
\newblock Manipulating {OpenFlow} link discovery packet forwarding for topology poisoning.
\newblock \url{https://zenodo.org/records/13292328}, 2024.

\bibitem{choi2017design}
Jin~Seek Choi, Sungtae Kang, and Young Lee.
\newblock Design and evaluation of a pcep-based topology discovery protocol for stateful pce.
\newblock {\em Optical Switching and Networking}, 26:39--47, 2017.

\bibitem{cisco_aci}
Cisco.
\newblock Open source used in cisco application policy infrastructure controller (apic) 1.2(1).
\newblock \url{https://www.cisco.com/c/dam/en/us/td/docs/switches/datacenter/aci/apic/sw/1-x/3rd-party/Cisco_ACI_Open_Source_1_1_3_v1_0.pdf}.

\bibitem{lldp}
Paul Congdon.
\newblock Link layer discovery protocol and mib v2.0.
\newblock \url{https://www.ieee802.org/1/files/public/docs2002/lldp-protocol-02.pdf}, 2002.
\newblock Accessed on 2023-06-21.

\bibitem{dhawan2015sphinx}
Mohan Dhawan, Rishabh Poddar, Kshiteej Mahajan, and Vijay Mann.
\newblock Sphinx: detecting security attacks in software-defined networks.
\newblock In {\em Ndss}, volume~15, pages 8--11, 2015.

\bibitem{ovs}
Linux Foundation.
\newblock Open vswitch manual.
\newblock \url{https://www.openvswitch.org/support/dist-docs/ovs-ofctl.8.html}.

\bibitem{opendaylight}
Linux Foundation.
\newblock Opendaylight.
\newblock \url{https://www.opendaylight.org/}.

\bibitem{onf_threat_analysis}
Open~Networking Foundation.
\newblock Threat analysis for the sdn architecture.
\newblock pages 1--21, 2016.

\bibitem{gao2022defense}
Yang Gao and Mingdi Xu.
\newblock Defense against software-defined network topology poisoning attacks.
\newblock {\em Tsinghua Science and Technology}, 28(1):39--46, 2022.

\bibitem{habib2022mitigating}
Sana Habib, Tiffany Bao, Yan Shoshitaishvili, and Adam Doup{\'e}.
\newblock Mitigating threats emerging from the interaction between sdn apps and sdn (configuration) datastore.
\newblock In {\em Proceedings of the 2022 on Cloud Computing Security Workshop}, pages 23--39, 2022.

\bibitem{hasan2017efficient}
Dana Hasan and Mohamed Othman.
\newblock Efficient topology discovery in software defined networks: Revisited.
\newblock {\em Procedia computer science}, 116:539--547, 2017.

\bibitem{hong2015poisoning}
Sungmin Hong, Lei Xu, Haopei Wang, and Guofei Gu.
\newblock Poisoning network visibility in software-defined networks: New attacks and countermeasures.
\newblock In {\em Ndss}, volume~15, pages 8--11, 2015.

\bibitem{hu2018multi}
Tao Hu, Zehua Guo, Peng Yi, Thar Baker, and Julong Lan.
\newblock Multi-controller based software-defined networking: A survey.
\newblock {\em IEEE access}, 6:15980--15996, 2018.

\bibitem{hu2021seapp}
Tao Hu, Zhen Zhang, Peng Yi, Dong Liang, Ziyong Li, Quan Ren, Yuxiang Hu, and Julong Lan.
\newblock Seapp: A secure application management framework based on rest api access control in sdn-enabled cloud environment.
\newblock {\em Journal of Parallel and Distributed Computing}, 147:108--123, 2021.

\bibitem{huang2020towards}
Xinli Huang, Peng Shi, Yufei Liu, and Fei Xu.
\newblock Towards trusted and efficient sdn topology discovery: A lightweight topology verification scheme.
\newblock {\em Computer Networks}, 170:107119, 2020.

\bibitem{hussain2021broadcast}
Mir~Wajahat Hussain, Soumen Moulik, and Diptendu~Sinha Roy.
\newblock A broadcast based link discovery scheme for minimizing messages in software defined networks.
\newblock In {\em 2021 IEEE Globecom Workshops (GC Wkshps)}, pages 1--6. IEEE, 2021.

\bibitem{jaume20202}
Daniel~A Jaume, Adri{\'a}n Pastine, and Victor~Nicolas Schv{\"o}llner.
\newblock 2-switch: transition and stability on graphs and forests.
\newblock {\em arXiv preprint arXiv:2004.11164}, 2020.

\bibitem{junos}
Juniper.
\newblock Junos space datasheet.
\newblock \url{https://www.juniper.net/us/en/products/sdn-and-orchestration/junos-space-datasheet.html}.

\bibitem{kazemian2013real}
Peyman Kazemian, Michael Chang, Hongyi Zeng, George Varghese, Nick McKeown, and Scott Whyte.
\newblock Real time network policy checking using header space analysis.
\newblock In {\em 10th USENIX Symposium on Networked Systems Design and Implementation (NSDI 13)}, pages 99--111, 2013.

\bibitem{khan2016topology}
Suleman Khan, Abdullah Gani, Ainuddin Wahid~Abdul Wahab, Mohsen Guizani, and Muhammad~Khurram Khan.
\newblock Topology discovery in software defined networks: Threats, taxonomy, and state-of-the-art.
\newblock {\em IEEE Communications Surveys \& Tutorials}, 19(1):303--324, 2016.

\bibitem{khurshid2012veriflow}
Ahmed Khurshid, Xuan Zou, Wenxuan Zhou, Matthew Caesar, and P.~Brighten Godfrey.
\newblock {VeriFlow}: Verifying {Network-Wide} invariants in real time.
\newblock In {\em 10th USENIX Symposium on Networked Systems Design and Implementation (NSDI 13)}, pages 15--27, Lombard, IL, April 2013. USENIX Association.

\bibitem{kim2023intender}
Jiwon Kim, Benjamin~E Ujcich, and Dave~Jing Tian.
\newblock Intender: Fuzzing $\{$Intent-Based$\}$ networking with $\{$Intent-State$\}$ transition guidance.
\newblock In {\em 32nd USENIX Security Symposium (USENIX Security 23)}, pages 4463--4480, 2023.

\bibitem{kreutz2014software}
Diego Kreutz, Fernando~MV Ramos, Paulo~Esteves Verissimo, Christian~Esteve Rothenberg, Siamak Azodolmolky, and Steve Uhlig.
\newblock Software-defined networking: A comprehensive survey.
\newblock {\em Proceedings of the IEEE}, 103(1):14--76, 2014.

\bibitem{lee2018indago}
Chanhee Lee, Changhoon Yoon, Seungwon Shin, and Sang~Kil Cha.
\newblock Indago: A new framework for detecting malicious sdn applications.
\newblock In {\em 2018 IEEE 26th International Conference on Network Protocols (ICNP)}, pages 220--230. IEEE, 2018.

\bibitem{Leiserson:1985}
Charles~E. Leiserson.
\newblock Fat-trees: Universal networks for hardware-efficient supercomputing.
\newblock {\em IEEE Transactions on Computers}, C-34(10), 1985.

\bibitem{li2014byzantine}
He~Li, Peng Li, Song Guo, and Shui Yu.
\newblock Byzantine-resilient secure software-defined networks with multiple controllers.
\newblock In {\em 2014 IEEE International Conference on Communications (ICC)}, pages 695--700. IEEE, 2014.

\bibitem{li2019application}
Tong Li, Jinqiang Chen, and Hongyong Fu.
\newblock Application scenarios based on sdn: an overview.
\newblock In {\em Journal of Physics: Conference Series}, volume 1187, page 052067. IOP Publishing, 2019.

\bibitem{maleh2023comprehensive}
Yassine Maleh, Youssef Qasmaoui, Khalid El~Gholami, Yassine Sadqi, and Soufyane Mounir.
\newblock A comprehensive survey on sdn security: threats, mitigations, and future directions.
\newblock {\em Journal of Reliable Intelligent Environments}, 9(2):201--239, 2023.

\bibitem{matsumoto2014fleet}
Stephanos Matsumoto, Samuel Hitz, and Adrian Perrig.
\newblock Fleet: Defending sdns from malicious administrators.
\newblock In {\em Proceedings of the third workshop on Hot topics in software defined networking}, pages 103--108, 2014.

\bibitem{pox}
Colin~Scott Murphy.
\newblock The pox network software platform.
\newblock \url{https://github.com/noxrepo/pox}, 2013.
\newblock Accessed on 2023-06-19.

\bibitem{nehra2019sldp}
Ajay Nehra, Meenakshi Tripathi, Manoj~Singh Gaur, Ramesh~Babu Battula, and Chhagan Lal.
\newblock Sldp: A secure and lightweight link discovery protocol for software defined networking.
\newblock {\em Computer Networks}, 150:102--116, 2019.

\bibitem{nehra2019tilak}
Ajay Nehra, Meenakshi Tripathi, Manoj~Singh Gaur, Ramesh~Babu Battula, and Chhagan Lal.
\newblock Tilak: A token-based prevention approach for topology discovery threats in sdn.
\newblock {\em International Journal of Communication Systems}, 32(17):e3781, 2019.

\bibitem{nencioni2017impact}
Gianfranco Nencioni, Bjarne~E Helvik, Andres~J Gonzalez, Poul~E Heegaard, and Andrzej Kamisinski.
\newblock Impact of sdn controllers deployment on network availability.
\newblock {\em arXiv preprint arXiv:1703.05595}, 2017.

\bibitem{samsung}
Samsung Newsroom.
\newblock Samsung expands its lineup of sdn solutions.
\newblock \url{https://news.samsung.com/global/samsung-expands-its-lineup-of-sdn-solutions}, 2021.

\bibitem{nguyen2017analysis}
Tri-Hai Nguyen and Myungsik Yoo.
\newblock Analysis of link discovery service attacks in sdn controller.
\newblock In {\em 2017 International Conference on Information Networking (ICOIN)}, pages 259--261. IEEE, 2017.

\bibitem{ochoa2015current}
Leonardo Ochoa~Aday, Cristina Cervell{\'o}~Pastor, and Adriana Fern{\'a}ndez~Fern{\'a}ndez.
\newblock Current trends of topology discovery in openflow-based software defined networks.
\newblock 2015.

\bibitem{ochoa2019etdp}
Leonardo Ochoa-Aday, Cristina Cervell{\'o}-Pastor, and Adriana Fern{\'a}ndez-Fern{\'a}ndez.
\newblock Etdp: Enhanced topology discovery protocol for software-defined networks.
\newblock {\em IEEE access}, 7:23471--23487, 2019.

\bibitem{topozoo}
University of~Adelaide.
\newblock Topology zoo.
\newblock \url{http://www.topology-zoo.org/dataset.html}.

\bibitem{pakzad2016efficient}
Farzaneh Pakzad, Marius Portmann, Wee~Lum Tan, and Jadwiga Indulska.
\newblock Efficient topology discovery in openflow-based software defined networks.
\newblock {\em Computer Communications}, 77:52--61, 2016.

\bibitem{papadimitriou2010web}
Panagiotis Papadimitriou, Ali Dasdan, and Hector Garcia-Molina.
\newblock Web graph similarity for anomaly detection.
\newblock {\em Journal of Internet Services and Applications}, 1:19--30, 2010.

\bibitem{pertsch2021accelerating}
Karl Pertsch, Youngwoon Lee, and Joseph Lim.
\newblock Accelerating reinforcement learning with learned skill priors.
\newblock In {\em Conference on robot learning}, pages 188--204. PMLR, 2021.

\bibitem{porras2012security}
Philip Porras, Seungwon Shin, Vinod Yegneswaran, Martin Fong, Mabry Tyson, and Guofei Gu.
\newblock A security enforcement kernel for openflow networks.
\newblock In {\em Proceedings of the first workshop on Hot topics in software defined networks}, pages 121--126, 2012.

\bibitem{qi2016intensive}
Chao Qi, Jiangxing Wu, Hongchao Hu, Guozhen Cheng, Wenyan Liu, Jianjian Ai, and Chao Yang.
\newblock An intensive security architecture with multi-controller for sdn.
\newblock In {\em 2016 ieee conference on computer communications workshops (infocom wkshps)}, pages 401--402. IEEE, 2016.

\bibitem{floodlight}
Ryan~Izard Qing~Wang, Geddings~Barrineau.
\newblock Floodlight sdn openflow controller.
\newblock \url{https://github.com/floodlight/floodlight}, 2016.
\newblock Accessed on 2023-06-19.

\bibitem{rojas2018tedp}
Elisa Rojas, Joaquin Alvarez-Horcajo, Isaias Martinez-Yelmo, Juan~A Carral, and Jose~M Arco.
\newblock Tedp: An enhanced topology discovery service for software-defined networking.
\newblock {\em IEEE Communications Letters}, 22(8):1540--1543, 2018.

\bibitem{ropke2018preventing}
Christian R{\"o}pke and Thosten Holz.
\newblock Preventing malicious sdn applications from hiding adverse network manipulations.
\newblock In {\em Proceedings of the 2018 Workshop on Security in Softwarized Networks: Prospects and Challenges}, pages 40--45, 2018.

\bibitem{russell2016artificial}
Stuart~J Russell and Peter Norvig.
\newblock {\em Artificial intelligence: a modern approach}.
\newblock Malaysia; Pearson Education Limited,, 2016.

\bibitem{shaghaghi2020software}
Arash Shaghaghi, Mohamed~Ali Kaafar, Rajkumar Buyya, and Sanjay Jha.
\newblock Software-defined network (sdn) data plane security: issues, solutions, and future directions.
\newblock {\em Handbook of Computer Networks and Cyber Security}, pages 341--387, 2020.

\bibitem{Gwardar18}
Arash Shaghaghi, Salil~S. Kanhere, Mohamed~Ali Kaafar, and Sanjay Jha.
\newblock Gwardar: Towards protecting a software-defined network from malicious network operating systems.
\newblock In {\em 2018 IEEE 17th International Symposium on Network Computing and Applications (NCA)}, pages 1--5, 2018.

\bibitem{skowyra2018effective}
Richard Skowyra, Lei Xu, Guofei Gu, Veer Dedhia, Thomas Hobson, Hamed Okhravi, and James Landry.
\newblock Effective topology tampering attacks and defenses in software-defined networks.
\newblock In {\em 2018 48th Annual IEEE/IFIP International Conference on Dependable Systems and Networks (DSN)}, pages 374--385. IEEE, 2018.

\bibitem{smyth2017detecting}
Dylan Smyth, Sean McSweeney, Donna O'Shea, and Victor Cionca.
\newblock Detecting link fabrication attacks in software-defined networks.
\newblock In {\em 2017 26th International Conference on Computer Communication and Networks (ICCCN)}, pages 1--8. IEEE, 2017.

\bibitem{tarnaras2015sdn}
George Tarnaras, Evangelos Haleplidis, and Spyros Denazis.
\newblock Sdn and forces based optimal network topology discovery.
\newblock In {\em Proceedings of the 2015 1st IEEE Conference on Network Softwarization (NetSoft)}, pages 1--6. IEEE, 2015.

\bibitem{tootoonchian2010hyperflow}
Amin Tootoonchian and Yashar Ganjali.
\newblock Hyperflow: A distributed control plane for openflow.
\newblock In {\em Proceedings of the 2010 internet network management conference on Research on enterprise networking}, volume~3, 2010.

\bibitem{openflow}
ONF TS-009.
\newblock Openflow switch specification version 1.3.2.
\newblock \url{https://opennetworking.org/wp-content/uploads/2014/10/openflow-spec-v1.3.2.pdf}.

\bibitem{tseng2018comprehensive}
Yuchia Tseng, Farid Na{\"\i}t-Abdesselam, and Ashfaq Khokhar.
\newblock A comprehensive 3-dimensional security analysis of a controller in software-defined networking.
\newblock {\em Security and Privacy}, 1(2):e21, 2018.

\bibitem{tseng2017controller}
Yuchia Tseng, Montida Pattaranantakul, Ruan He, Zonghua Zhang, and Farid Na{\"\i}t-Abdesselam.
\newblock Controller dac: Securing sdn controller with dynamic access control.
\newblock In {\em 2017 IEEE International Conference on Communications (ICC)}, pages 1--6. IEEE, 2017.

\bibitem{ujcich2018cross}
Benjamin~E Ujcich, Samuel Jero, Anne Edmundson, Qi~Wang, Richard Skowyra, James Landry, Adam Bates, William~H Sanders, Cristina Nita-Rotaru, and Hamed Okhravi.
\newblock Cross-app poisoning in software-defined networking.
\newblock In {\em Proceedings of the 2018 ACM SIGSAC conference on computer and communications security}, pages 648--663, 2018.

\bibitem{ujcich2021causal}
Benjamin~E Ujcich, Samuel Jero, Richard Skowyra, Adam Bates, William~H Sanders, and Hamed Okhravi.
\newblock Causal analysis for $\{$Software-Defined$\}$ networking attacks.
\newblock In {\em 30th USENIX Security Symposium (USENIX Security 21)}, pages 3183--3200, 2021.

\bibitem{wazirali2021sdn}
Raniyah Wazirali, Rami Ahmad, and Suheib Alhiyari.
\newblock Sdn-openflow topology discovery: an overview of performance issues.
\newblock {\em Applied Sciences}, 11(15):6999, 2021.

\bibitem{ONOS_telecom}
Alan Weissberger.
\newblock Comcast: Onf trellis software is in production together with l2/l3 white box switches.
\newblock \url{https://techblog.comsoc.org/2019/09/14/comcast-puts-onf-trellis-software-into-production/}.

\bibitem{pica8_manual2}
Tracy Yang.
\newblock Picos 4.4.4 configuration guide (special release): Ecmp select group.
\newblock \url{https://pica8-fs.atlassian.net/wiki/spaces/PicOS4443beta/pages/115898390/Ecmp+Select+Group}.

\bibitem{pica8_manual1}
Tracy Yang.
\newblock Picos 4.4.4 configuration guide (special release): ovs-ofctl add-flow <bridge> <flow>.
\newblock \url{https://pica8-fs.atlassian.net/wiki/spaces/PicOS4443beta/pages/115900326/ovs-ofctl+add-flow+bridge+flow}.

\bibitem{blackhat}
Ch~Yoon and S~Lee.
\newblock Attacking sdn infrastructure: Are we ready for the next-gen networking?
\newblock {\em BlackHat-USA-2016}, pages 17--18, 2016.

\bibitem{zhang2018survey}
Yuan Zhang, Lin Cui, Wei Wang, and Yuxiang Zhang.
\newblock A survey on software defined networking with multiple controllers.
\newblock {\em Journal of Network and Computer Applications}, 103:101--118, 2018.

\bibitem{zhao2018esld}
Xin Zhao, Lin Yao, and Guowei Wu.
\newblock Esld: An efficient and secure link discovery scheme for software-defined networking.
\newblock {\em International Journal of Communication Systems}, 31(10):e3552, 2018.

\bibitem{zhou2018sdn}
Haifeng Zhou, Chunming Wu, Chengyu Yang, Pengfei Wang, Qi~Yang, Zhouhao Lu, and Qiumei Cheng.
\newblock Sdn-rdcd: A real-time and reliable method for detecting compromised sdn devices.
\newblock {\em IEEE/ACM transactions on networking}, 26(5):2048--2061, 2018.

\end{thebibliography}

\appendix

\section{Poisonous Flow Entry Computation}\label{sec:appendix_pois_flow_entry}

Algorithm \ref{alg:pois_link} shows a VLAN-based precise link manipulation algorithm that uses the \kw{ether-src}-distinguishable VLAN poisonous flow entries described in Section \ref{sec:approach}. Information about the link is collected in Lines 1--4. There may be multiple paths that can be used to build $Ax\rightarrow yB$, but the shortest path between $A$ and $B$ is not necessarily the best. Lines 5--11 determine which path will minimize the number of required poisonous flow entries.

\begin{figure}[t]
\vspace{-1\baselineskip}
\centering
\subfigure[Loopback Method]{\label{fig:loopback} \includegraphics[width=0.2\textwidth]{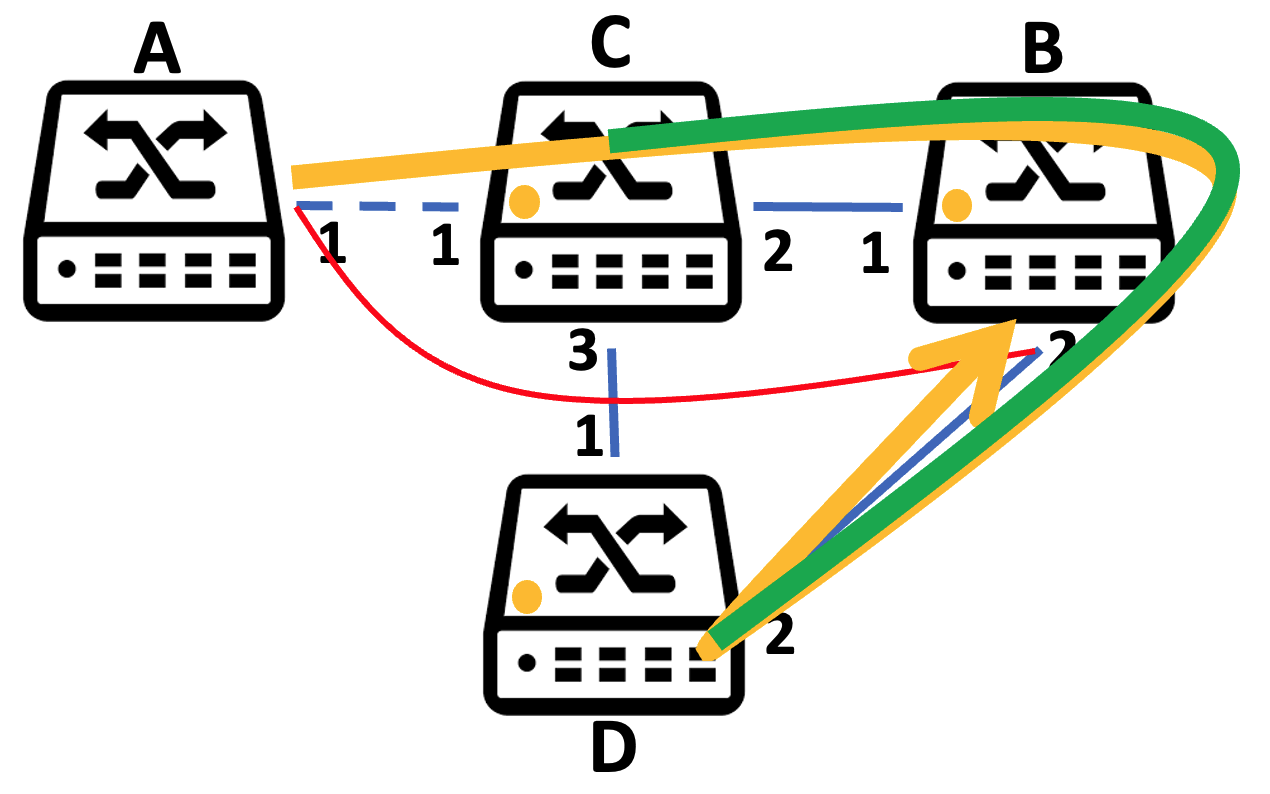}}
\hfill
\subfigure[No-loop Method]{\label{fig:no_loop} \includegraphics[width=0.22\textwidth]{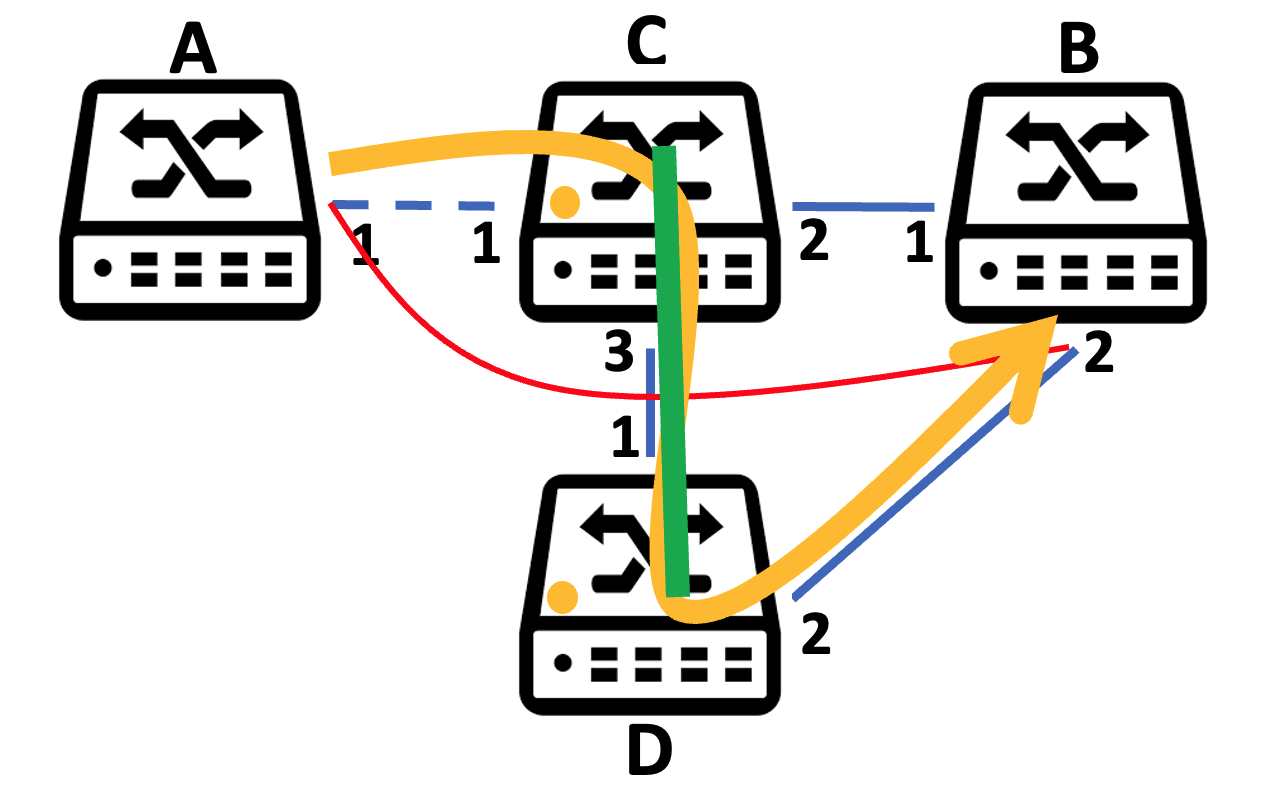}}%
\hfill
\caption{Path selection of Loopback and No-loop methods}
\label{fig:sele_rou}
\vspace{-1\baselineskip}
\end{figure}

While we use Dijkstra's shortest path routing algorithm to compute the poisoning path, additional poisonous flow entry setups may be required depending on the target deceptive link. In certain situations, it is necessary to forward the packet back to its received port. To guarantee that we can fabricate links that terminate on the proper ports of the switches, there are two methods to create poisonous paths. The first is a direct method that results in a \textit{no-loop} path.  In circumstances in which the direct method cannot terminate the poisonous link on the correct port, we must use an indirect method that results in a \textit{loopback} poisonous path.  Suppose the deceptive link is $A1\rightarrow 2B$ shown in red in Figure \ref{fig:sele_rou}. The discovery packet is initiated by the controller and sent to $A$ within a \kw{packet-out} message. $A$ sends the discovery packet to the first hop $C$ (which is $src$ in Line 3). 
This process cannot be manipulated by \projname{} because the framework does not tamper with the \kw{packet-out} message; therefore, $C$ is the source of our shortest path calculation to find a path from $A1$ to $2B$.  However, Dijkstra's algorithm is not port number sensitive. As a result, it returns a solution of the shortest path ($C\rightarrow B$) with length 1, but as shown, the direct path does not terminate on $(B, Port:2)$. In this case we must resort to the indirect method to find the {loopback} path.   
To accomplish this, the discovery packet must be sent out through $(B, Port:2)$ and have $(D, Port:2)$ send it back so it can be received at $(B, Port:2)$ as shown in Figure~\ref{fig:loopback}.
We need the loopback in certain cases because if we are targeting $(B, Port:2)$ the LLDP packet must be received on a port from a link \textit{outside} the switch to be checked against the switch's flow table, and then passed to the controller. 
\begin{algorithm}[t]
\caption{Algorithm of Precise Link Manipulation}
\begin{algorithmic}[1]\label{alg:pois_link}
\footnotesize
    \REQUIRE $T$: Topology, $N$: Nodes, $Ax\rightarrow yB$: fake link, $vlan\_id$: VLAN ID
    \ENSURE $Ax\rightarrow yB$ is set up 
    \STATE $flag_{loopback} \coloneqq False$
    \STATE $src\_eth \coloneqq $  Get\_MAC$(N, Ax)$
    \STATE $sub\_src \coloneqq $ Find\_Nei$(T, Ax)$\COMMENT{The first hop after the $Ax$}
    \STATE $sub\_dst \coloneqq $ Find\_Nei$(T, yB)$\COMMENT{The last hop before the $yB$}
    \IF{$|$Dij$(T, sub\_src, sub\_dst)| \leq |$Dij$(T, sub\_src, B)|$ }
        \STATE $sub\_path \coloneqq $ Dij$(T, sub\_src, sub\_dst)$
        \STATE $path \coloneqq \{sub\_path, B\}$
    \ELSE
        \STATE $path \coloneqq$ Dij$(T, sub\_src, B)$
        \STATE $flag_{loopback} \coloneqq True$
    \ENDIF
    \STATE{$match_{simple}\coloneqq [ETH\_SRC: src\_eth]$}
    \STATE{$match_{inport}(in\_port) \coloneqq [ETH\_SRC:src\_eth, IN\_PORT: in\_port]$}
    \STATE{$match_{Vid}\coloneqq [VLAN\_VID:vlan\_id]$}
    \STATE{$actions_{simple}(out\_port)\coloneqq [OUTPUT: out\_port]$}
    \STATE{$actions_{pushV}(out\_port)\coloneqq [PUSH\_VLAN:vlan\_id, OUTPUT: out\_port]$}
    \STATE{$actions_{popV}(out\_port)\coloneqq[POP\_VLAN, OUTPUT: out\_port]$}
    \FOR{$i = 0$ to $|path|-1$}
        \STATE{$out\_port \coloneqq$ Get\_LPort$(T, path[i], path[i+1])$}
        \IF{$i== 0$ and $|path|==2$ and not $flag_{loopback}$}
            \STATE{$\underline{E_{hop}} \coloneqq$ Send\_PoisE$(path[i], match_{simple}, actions_{simple}(out\_port))$}
        \ELSIF{$i==0$} 
            \IF{$path[i+1]$ is $A$}
                \STATE{$out\_port = IN\_PORT$}
            \ENDIF
            \STATE{$\underline{E_{VStart}} \coloneqq$ Send\_PoisE$(path[i], match_{simple}, actions_{pushV}(out\_port))$}
        \ELSIF{$i == path[|path|-2]$ and not $flag_{loopback}$} 
            \STATE{$\underline{E_{VEnd}}\coloneqq$ Send\_PoisE$(path[i], match_{Vid}, actions_{popV}(out\_port))$}         
        \ELSE 
            \STATE{$\underline{E_{VBody}} \coloneqq$ Send\_PoisE$(path[i], match_{Vid}, actions_{simple}(out\_port))$}
        \ENDIF
    \ENDFOR
    \IF{$flag_{loopback}$}
        \IF{$|path|>1$}
            \STATE{$\underline{E_{VBody}}\coloneqq$ Send\_PoisE$(B, match_{Vid}, actions_{simple}(y))$}
        \ELSE
            \STATE{$out\_port \coloneqq$  Get\_RPort$(T,A, B)$}
            \STATE{$\underline{E_{VStart}}\coloneqq$ Send\_PoisE$(B, match_{inport}(out\_port), actions_{pushV}(y))$}
        \ENDIF
        \STATE{$\underline{E_{VEnd}}\coloneqq$ Send\_PoisE$(sub\_dst, match_{Vid}, actions_{popV}(IN\_PORT))$}
    \ENDIF
\end{algorithmic}
    
\end{algorithm}

While the loopback method can 
create deceptive links successfully, it may require more flow entries than needed.  In the example, three flow entries are configured on three switches ($C$, $B$, $D$). By contrast, the no-loop method shown in Figure~\ref{fig:no_loop} only requires two entries on $C$ and $D$. When this is the case, the algorithm chooses the no-loop method (lines 5--7); when it is not, it chooses the loopback method (lines 8--11). 

The poisonous flow entries computed in Lines 12--41 forward the discovery packet along the poisoning path. There are four types of flow entries. The $E_{VStart}$, $E_{VEnd}$, and $E_{VBody}$ flow entries are for building the VLAN poisonous paths (the green paths in Figure~\ref{fig:sele_rou}) as described in \S\ref{sec:vlanentries}. $E_{VStart}$ adds a VLAN tag to an LLDP packet and sends it out, $E_{VEnd}$ matches the VLAN tag, removes the tag and forwards the original packet, and $E_{VBody}$ simply matches the tag and forwards the packet. $E_{hop}$ (lines 20--21) is the flow entry for a single hop when the path is too short to build a VLAN tunnel and only a single hop is needed.
%
In lines 12--17, we compose the matches and actions for $E_{hop}$, $E_{VStart}$, $E_{VEnd}$, and $E_{VBody}$ based on the vanilla and VLAN poisonous flow entry design discussed in \S\ref{sec:approach}. In lines 18--41, we relay the discovery packet as appropriate to either set up a VLAN tunnel with or without loopback or relay the packet a single hop if that is sufficient to build the deceptive link. 

There are two critical points when composing the VLAN flow entries. First, the loopback can happen at both the beginning (lines 23--26) (called head loopback) and the end (lines 33--41) (called tail loopback) of the VLAN tunnel.  Second, nodes that act as both the initiator of a tail loopback, and the beginning node for a VLAN tunnel must be carefully crafted to avoid flow entry conflict (lines 36--39). In such instances, the same discovery packets will be received twice at the node, but require different actions; therefore, we need to distinguish at what stage of the forwarding sequence the packet is in by matching against the \kw{in-port} and \kw{ether-src} fields.

\section{Gap-Patching Flow Entry Computation}\label{sec:appendix_gap_patching}

The logic of the reactive gap patching is simple: when a sensitive flow entry is detected such that it forwards packets through the endpoint of a fabricated link (line 2), the Gap Patching module is triggered to set gap-patching flow entries on $src$ (the next hop of the endpoint). The port identifiers and VLAN ID can be extracted from the switches to determine if the gap is a single hop or longer.  
If the gap is a single hop, the patching is a simple action of outputting the packet to \kw{out\_port\_num}.  If the gap is more than one hop, the action is to push the VLAN ID. We create the matches differently to avoid detection by Sphinx~\cite{dhawan2015sphinx}. The reactive gap patching algorithm is shown in Algorithm \ref{alg:gap_patching}.

In a controller cluster scenario, the malicious controller gets any data store update in the cluster, including the flow entries configured by the lead controller, due to the data store consistency mechanism. As a result, \projname{} gets notifications of flow entry installation for free. The gap-patching delay is essentially influenced by the event update delay in the cluster and flow entry setup delay. In the SDN application scenario, the malicious application has to pull the data plane flow table changes periodically. The gap-patching delay is mainly influenced by the frequency that the controller queries the data plane with \kw{multipart-request flow} and the frequency the malicious application requests it from the controller, which varies depending on the implementation.
\begin{algorithm}[t]
\caption{Algorithm of Gap-Patching}
\begin{algorithmic}[1]\label{alg:gap_patching}
\footnotesize
    \REQUIRE $ReVid\_map$: \{Key: Right endpoints of fake links, Value: VLAN\_IDs\}, $ReFe\_map$: \{Key: Right endpoints of fake links, Value: flow entry information for next hop configuration\}, $T$: Topology, $[ep]$: endpoints of fake links
    \ENSURE All routings work
    \WHILE{Monitoring on $[ep]$}
        \IF{Exists an entry $E$ forwarding packets through $ep\in[ep]$}
        \STATE{$(in\_port, out\_port) \coloneqq ReFe\_map.get(ep)$}
        \STATE{$match \coloneqq [extract\_match(E), IN\_PORT:in\_port]$}
        \STATE{$vlan\_id \coloneqq ReVid\_map.get(ep)$}
        \IF{$vlan\_id$}
            \STATE{$actions \coloneqq [PUSH\_VLAN: vlan\_id, OUTPUT: out\_port]$}
        \ELSE
            \STATE{$actions \coloneqq [OUTPUT: out\_port]$}
        \ENDIF
        \STATE{$sub\_src \coloneqq Find\_Nei(T,ep)$}
        \STATE{$E_{mal} \coloneqq Send\_MalE(sub\_src, match, actions)$}
        \ENDIF
    \ENDWHILE
\end{algorithmic}
\end{algorithm}

\section{Controller Impersonation Attack on Controller Cluster}\label{sec:eval_impersonation}
We successfully carried out impersonation attacks on both ONOS and OpenDaylight clusters. In both cases, we formed three-node clusters.  In each case, the third controller ONOS/ODL-3 is in passive replication mode without a direct OpenFlow connection to the mininet. First, we shut down the ONOS/ODL-3 to mimic node failure. Second, we create ONOS/ODL-mal with the same IP address as ONOS/ODL-3 and configure an incomplete cluster configuration file. The incomplete configuration file only needs the offline controller's IP address and any one of the other controller's IP addresses in the cluster. Last, The ONOS/MAL-mal joins the cluster anyway after rebooting to bring the cluster configuration into effect, disclosing the ONOS/OpenDaylight cluster vulnerability.

The critical aspect of this attack lies in the flaw within the cluster rejoining implementation. Despite the absence of cluster-joining authentication, a candidate controller should only be accepted if it possesses the addresses of all cluster members properly configured. However, recent versions of ONOS and ODL clusters have been found to accept incomplete configurations, including only two members' addresses. Suppose there are $n$ machines within the subnet, and the offline time of ONOS/ODL-3 is $t_{offline}$. The time required for brute-force probing of cluster-joining messages with all $p(n,2)$ permutations of addresses is $t_{probe}$. If $t_{probe}<t_{offline}$, the controller impersonation attack can succeed. Under normal circumstances, $p(n,m)$ permutations are necessary to be probed to join the cluster, where $m$ is the number of controllers in the cluster. The difference between $O(n^2)$ and $O(n^m)$ is substantial, especially considering that $m>3$ is mandatory for forming a cluster.

\end{document}